\documentclass[journal,comsoc]{IEEEtran}
\usepackage{amsmath}
\usepackage{tabularx,multirow}
\usepackage{graphicx,subfigure,comment}
\usepackage[table]{xcolor}
\usepackage{url} 
\usepackage{psfrag,dsfont}
\usepackage{cite}
\usepackage{siunitx}
\usepackage{tabularx}

\usepackage[algoruled,medskip,dontprintsemicolon,linesnumbered,Algorithm]{algorithm}
\usepackage[noend]{algpseudocode}

\newtheorem{theorem}{Theorem}

\newcommand{\Fig}[1]{Fig.~\ref{fig:#1}}
\newcommand{\Sec}[1]{Sec.~\ref{sec:#1}}
\newcommand{\Tab}[1]{Tab.~\ref{tab:#1}}
\newcommand{\Eq}[1]{(\ref{eq:#1})}
\newcommand{\Alg}[1]{Alg.~\ref{alg:#1}}
\newcommand{\Line}[1]{Line~\ref{line:#1}}

\newcommand{\Thm}[1]{Theorem~\ref{thm:#1}}

\newtheorem{property}{Property}

\newcommand{\Vc}{\mathcal{V}}
\newcommand{\Ec}{\mathcal{E}}

\newcommand{\Sc}{\mathcal{S}}
\newcommand{\Cc}{\mathcal{C}}
\newcommand{\Lc}{\mathcal{L}}

\newcommand{\Pc}{\mathcal{P}}
\newcommand{\Wc}{\mathcal{W}}

\definecolor{mylightblue}{HTML}{1E90FF}
\definecolor{mydarkblue}{HTML}{000087}
\definecolor{mygreen}{HTML}{008A00}
\definecolor{myorange}{HTML}{FF5300}
\definecolor{mypink}{HTML}{A020F0}

\begin{document}

\title{
An Optimization-enhanced MANO \\ for Energy-efficient 5G Networks
} 

\author{
Francesco~Malandrino,~\IEEEmembership{Senior~Member,~IEEE,}
Carla-Fabiana~Chiasserini,~\IEEEmembership{Fellow,~IEEE,}
Claudio~Casetti,~\IEEEmembership{Senior~Member,~IEEE,}
Giada~Landi, Marco~Capitani
\IEEEcompsocitemizethanks{\IEEEcompsocthanksitem F.~Malandrino and C.-F.~Chiasserini are with CNR-IEIIT, Italy and Politecnico di Torino, Italy. C.~Casetti is with Politecnico di Torino, Italy. G.~Landi and M.~Capitani are with Nextworks~s.r.l., Pisa, Italy.
\IEEEcompsocthanksitem This work has been performed in the framework of the European Union's Horizon 2020 project 5G-EVE co-funded by the EU under grant agreement No 815074. The views expressed are those of the authors and do not necessarily represent the project. The Commission is not liable for any use that may be made of any of the information contained therein.
}
}

\maketitle

\begin{abstract}
5G network nodes, fronthaul and backhaul alike, will have both forwarding and computational capabilities. This makes energy-efficient network management more challenging, as decisions such as activating or deactivating a node impact on both the ability of the network to route traffic and the amount of processing it can perform. To this end, we formulate an optimization problem accounting for the main features of 5G nodes and the traffic they serve, allowing {\em joint} decisions about (i) the nodes to activate, (ii) the network functions they run, and (iii) the traffic routing. Our optimization module is integrated within the management and orchestration framework of 5G, thus enabling swift and high-quality decisions. We test our scheme with both a real-world testbed based on OpenStack and OpenDaylight, and a large-scale emulated network whose topology and traffic come from a real-world mobile operator, finding it to consistently outperform state-of-the art alternatives and closely match the optimum.
\end{abstract}

\section{Introduction}

Among the disruptive changes introduced by 5G networks, a major one is represented by the blurring of the distinction between forwarding equipment (e.g., switches) and computational facilities (e.g., servers). Indeed, backhaul and fronthaul nodes of 5G networks (hereinafter referred to as B/F nodes) will be endowed with computational, storage, and networking capabilities, allowing them to run any {\em virtual network function (VNF)}, from switches to video transcoders. VNFs are subsequently combined into {\em VNF graphs}, which define the services made available to higher network layers or third parties (e.g., {\em vertical} industries operating in the automotive, e-health, or media domain).

In this context, the entities of the {\em management and orchestration} (MANO) framework are in charge of making and implementing a set of complex decisions, including (i) activation of B/F nodes, so as to minimize the energy they consume, hence the costs for the operator; (ii) which VNF instances each B/F node shall run, in order to honor the delay constraints associated with the supported services; (iii) how traffic should be routed through the links connecting the B/F nodes. In traditional networks, these decisions could be made separately, owing to the fact that they concern different sets of equipment. 
Network design problems took as an input a static traffic matrix and, similarly, server placement problems assumed a known and immutable network topology. 
In 5G, on the other hand, decisions -- e.g., activating or deactivating a B/F node -- affect both the forwarding and computational capabilities of the network. It follows that traditional approaches may be ineffective, and often not even viable.

\begin{figure}[t]
\centering
\includegraphics[width=.7\columnwidth]{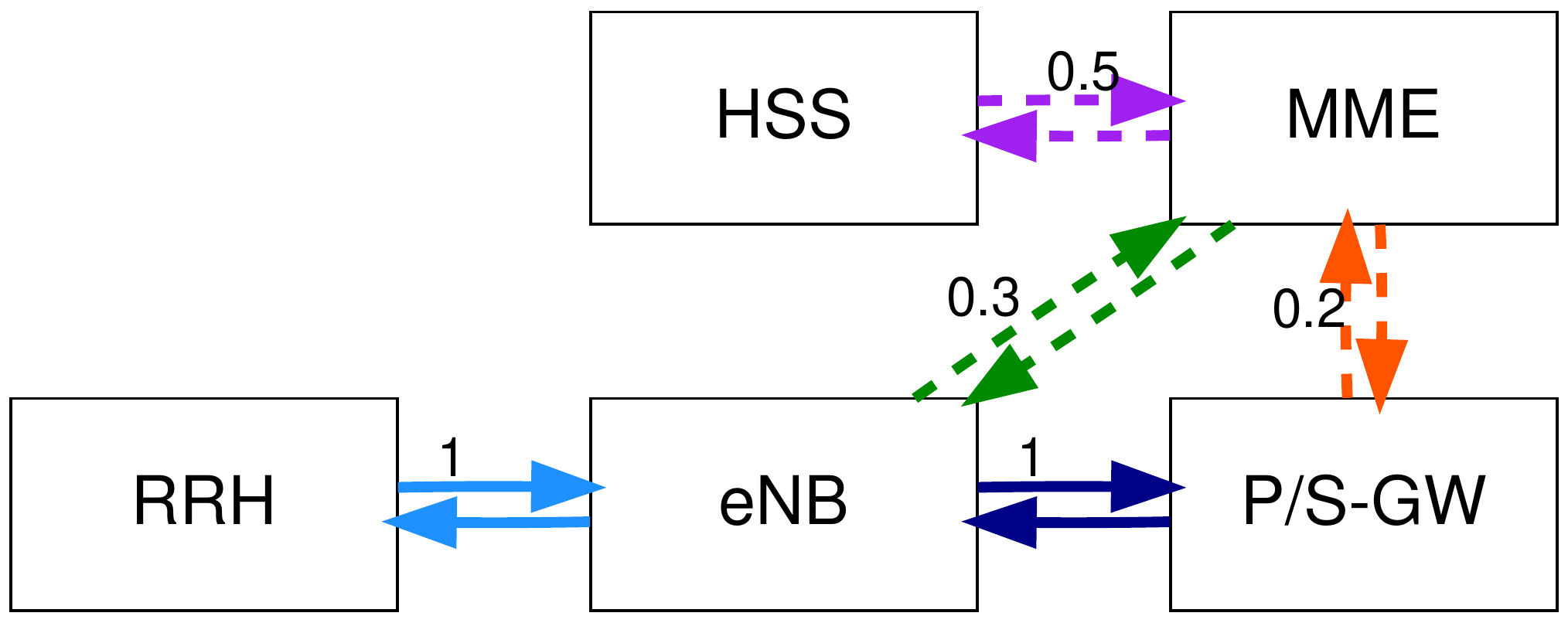}
\caption{
    Logical graph for vEPC. Solid lines correspond to user traffic, dashed lines to control traffic.
    \label{fig:logical}
} 
\end{figure}

The nature of 5G traffic further exacerbates this challenge. Indeed, as exemplified in \Fig{logical}, traffic flows in 5G need to traverse a {\em logical} graph whose vertices are VNFs; such graphs can have arbitrary complexity and are not restricted to being chains or directed acyclic graphs (DAGs). The task of the MANO entities can be described as matching such a logical graph with a {\em physical} graph whose vertices are B/F nodes and whose edges are the links, be them physical or virtual, that connect them. Such a matching must account for the fact that the quantity of traffic does {\em not} remain constant across processing steps (i.e., VNFs); in other words, the usual flow conservation laws do not hold.

\Fig{logical}, depicting the VNFs composing the virtual Evolved Packet Core (vEPC), depicts a typical example of this situation. Data-plane traffic flows from the remote radio head (RRH) to the eNodeB (eNB), and thence to the Packet/Service Gateway (P/S-GW). However, such a flow generate {\em additional} control-plane flows, e.g., going from the eNB to the Home Subscriber Server (HSS) through the Mobility Management Entity (MME). Even data traffic may not remain constant: as an example, firewalls and deep packet inspection (DPI) VNFs can drop some flows, thereby decreasing the network traffic from a processing step to the next.

Along with these challenges, the hybrid nature of 5G network nodes and their ability to be programmed through software results in significant opportunities, including the possibility to {\em optimize the management} of the network. Indeed, optimization is traditionally used in network design, but it is regarded to as too complex for their real-time management. In our work, we depart from this vision and {\em integrate} optimization within the MANO framework, thereby allowing its entities to make and implement high-quality and real-time decisions.

The main contributions of our paper are as follows:
\begin{itemize}
    \item a {\em model}, capturing the unique features of 5G network nodes (e.g., their hybrid nature) and of the traffic they serve (e.g., no flow conservation);
    \item a {\em problem formulation}, allowing us to make joint decisions on (i) B/F node activation, (ii) number and placement of the VNF instances, and (iii) traffic routing;
    \item a {\em solution concept}, named OptiLoop, predicated on integrating optimization {\em in the loop} of the decisions made by MANO entities, namely the NFV orchestrator (NFVO);
    \item two {\em implementations} of OptiLoop, one within a real-world testbed based on OpenStack and OpenDaylight, and one within a larger-scale network emulated in Mininet.
\end{itemize}

The remainder of this paper is organized as follows. We review related work in \Sec{relwork}, and explain how our own work fits within the management and orchestration (MANO) framework proposed by ETSI in \Sec{mano}. Next, we present our system model and problem formulation in \Sec{model}, and detail the OptiLoop solution strategy in \Sec{heur}. We then describe our testbeds' architecture, reference scenario and benchmarks in \Sec{scenario}, present numerical results in \Sec{results}, and conclude the paper in \Sec{conclusion}.

\section{Related work}
\label{sec:relwork}

Many works on VNF placement and traffic routing, including~\cite{swfan16_coopetition,infocom15_optimal,access16_joint}, take the approach of {\em matching} VNF and physical topology graphs, also proposing efficient solution strategies for the ensuing mixed-integer linear programming (MILP) problems. The optimization objectives are: minimizing network usage in~\cite{swfan16_coopetition}, minimizing VNF deployment cost in~\cite{infocom15_optimal}, minimizing CAPEX and OPEX in~\cite{access16_joint}. The later work~\cite{infocom16_deploying} takes an iterative approach, making VNF placement and routing decisions when a request arrives. \cite{tc16_delayaware}~takes the VNF placement as given and focuses on scheduling and routing.

Other works focus on the interaction between mobile operators and third parties using their services. As an example, \cite{jsac17_buysell}~considers a market where operators bid to serve incoming demands. Among energy-aware works, \cite{cloudnet16_energyaware} seeks to optimize VNF placement and job scheduling in order to minimize energy consumption. However, the algorithm presented in~\cite{cloudnet16_energyaware}  optimizes the server utilization but neglects the energy consumed by network elements such as B/F nodes.

Among the services that can be provided through SDN/NFV-based networks, a prominent example is the EPC. As suggested by the survey in~\cite{epc_survey}, ILP and MILP are the most popular modeling tools, and heuristic algorithms the most popular solution strategy. A common theme~\cite{epc_split,globecom15_bearer,qos_split} is splitting EPC elements, e.g., the Packet Gateway (P-GW) and Service Gateway (S-GW), into separate sub-elements, one dealing with control traffic and the other with user traffic.  \cite{vepc_archi}  finds that such an approach reduces the total cost of ownership. Interestingly, other works, e.g.,~\cite{tvt_mme,openair}, take the opposite approach and merge P-GW and S-GW in a single entity (the P/S-GW). \cite{tvt_mme} focuses on the MME and proposes to implement it through four separate VNFs, whose number can vary so as to accommodate traffic fluctuations. Closer to our own effort is the recent work in~\cite{comsnets17_placement}, which studies the problem of placing the VNFs implementing the main EPC network functions -- S-GW, P-GW and MME -- across the available physical machines, subject to limits on their power and link capacity. A preliminary version~\cite{noi-wowmom18} of this work addressed the same problem, albeit in simpler scenarios and with a more limited scope.

\subsection{Novelty}

Our approach is novel with respect to the above works in several important ways:
\begin{enumerate}
    \item first and foremost, the scope of our work: we jointly account for (i) the number and placement of VNF instances,  
(ii) traffic routing,  and
(iii) 
network management, e.g., activating/deactivating B/F nodes and links; 
    \item at the modeling level: accounting for the complexity of 5G traffic, with requests that originate at a network endpoint and traverse multiple VNFs, triggering additional requests as they do so (hence the quantity of traffic changes across processing steps);
    \item as far as objectives are concerned: adopting energy-saving as our priority and using detailed and realistic energy models, instead of  proxy metrics as in~\cite{cloudnet16_energyaware};
    \item from a solution strategy viewpoint: optimizing in the loop, i.e., using optimization as a tool rather than a mere analysis technique;
\item at implementation level: validating and testing our approach through a testbed based on OpenDaylight and OpenStack. 
\end{enumerate}


\section{OptiLoop and the ETSI MANO framework}
\label{sec:mano}

The management and orchestration (MANO) framework, standardized by ETSI
in~\cite{etsimano}, includes a set of decision-making entities ({\em
functional blocks}) in charge of managing NFV-based networks, along with
the interfaces ({\em reference points}) between them. The high-level
goal of the framework is to map the key performance indicators (KPIs) chosen by the verticals, e.g.,
maximum end-to-end latency, into decisions concerning the network
resources, e.g., the activation/deactivation of (virtual) servers and
the placement of VNFs therein. In the following, we present a short
overview of the framework and then, in \Sec{mano-r-us}, discuss the
relationship between the NFV orchestrator, one of the most important
MANO entities, and OptiLoop.


\begin{figure*}
\centering
\includegraphics[width=1.4\columnwidth]{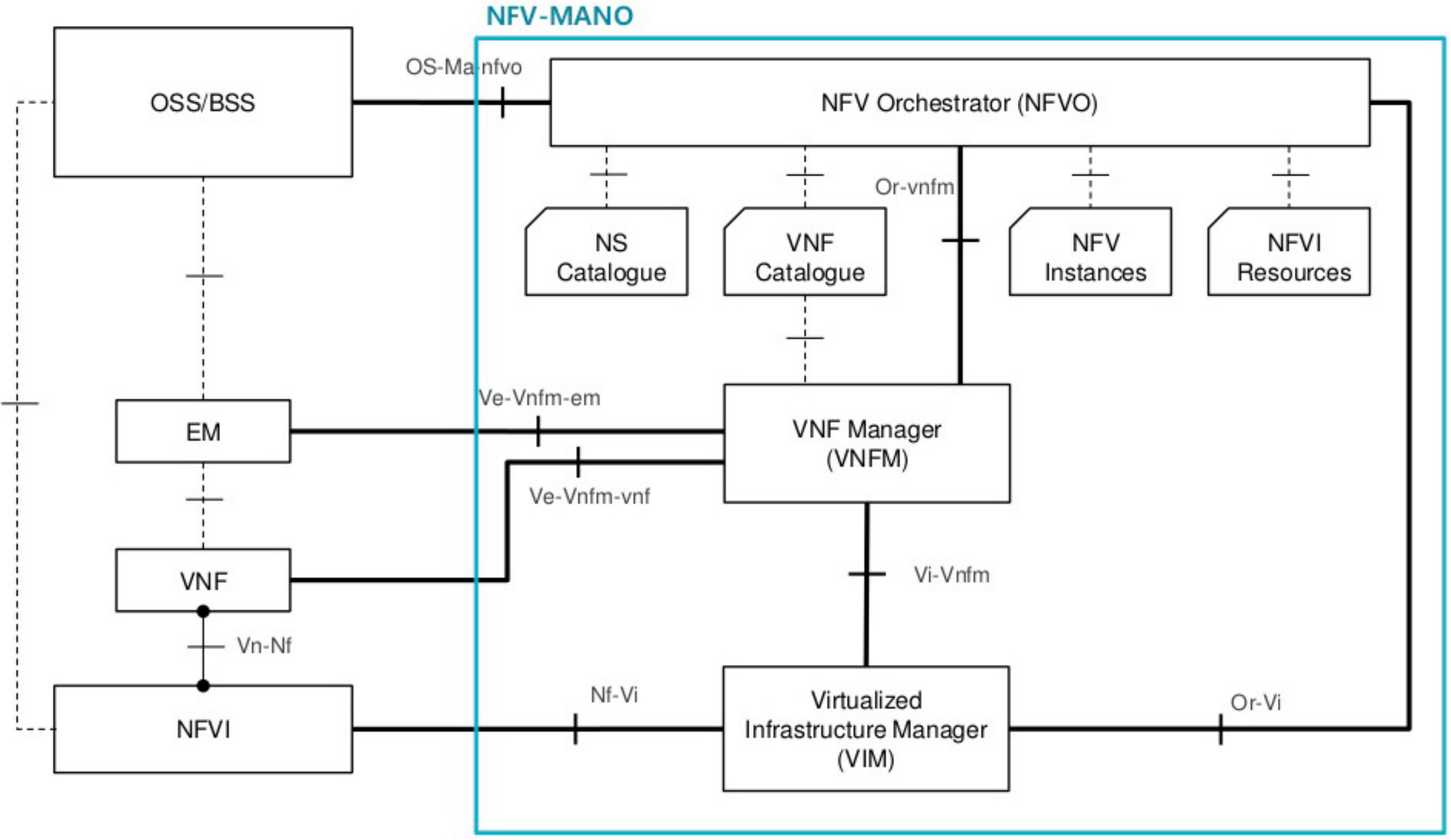}
\caption{
    The NFV-MANO architectural framework. Source:~\cite{etsimano}
    \label{fig:mano}
} 
\end{figure*}

\Fig{mano}, taken from~\cite{etsimano}, shows the decision-making
entities of the MANO framework (within the blue area), along with the
non-MANO entities they interact with. OSS/BSS (Operation and Business
Support Services), at the top-left corner, are the contact point between
verticals and mobile operators: they collect the vertical requirements,
expressed through end-to-end KPIs, and convey them, through the
Os-Ma-nfvo reference point, to the NFV Orchestrator (NFVO). 
The NFVO itself is arguably one of the most important entities of the MANO framework, and is in charge of the orchestration decisions. Specifically, given the vertical requirements and the state of the network infrastructure, the NFVO determines:
\begin{itemize}
    \item how many instances of each VNF to deploy;
    \item where in the network infrastructure they shall run;
    \item the features of the virtual network connecting the VNF instances, e.g., the bandwidth of the links to traverse between them.
\end{itemize}

Through the Or-vnfm interface, these decisions are sent to the VNFM (VNF
Manager), which takes care of instantiating the VNFs, requesting to the
VIM (Virtual Infrastructure Manager) the needed resources, e.g., virtual
machines or virtual links. The VIM, in turn, interacts with the NFVI
(NFV Infrastructure), which includes the servers running the VNFs and
the hypervisors managing them. The VNFM also communicates with a
non-MANO entity called EM (Element Manager), in charge of FCAPS (Fault,
Configuration, Accounting, Performance and Security) management, in
order to configure the VNFs or collect/monitor KPIs from them.

\subsection{The NFVO: input, output, and decisions}
\label{sec:mano-r-us}

The NFVO is in charge of most of the orchestration
tasks in the MANO framework. Owing to its importance, in the following
we detail the decisions it is in charge of, along with the input
information it has access to; such pieces of information correspond,
respectively, to the output and input of OptiLoop.

The main input data used by the NFVO is the NSD (Network Service
Descriptor), a data structure defined in~\cite[Sec.~6.2.1]{etsimano}.
NSDs contain a graph description of the VNFs needed by each service,
called VNFFG (VNF Forwarding Graph)~\cite[Sec.~6.5.1]{etsimano} along
with {\em deployment flavor} information, including the maximum latency
acceptable for each service\cite[Sec.~6.2.1.3]{etsimano}. Furthermore,
the NFVO has access to information on the network infrastructure, e.g.,
the state and capabilities of network and computing resources available
at the NFV infrastructure, including details about the connectivity
among the servers where the VNFs will be allocated.

Using all the above, the NFVO makes decisions about:
\begin{itemize}
    \item the status of network infrastructure elements, e.g., servers;
    \item VNF {\em lifecycle management}~\cite[Sec.~7.2]{etsimano} about the VNFs, including the host they run at;
    \item {\em routing}, accounting for the capacity and delay of virtual links.
\end{itemize}
Such decisions will correspond to decision variables in our system model, as detailed next. 

\section{System model}
\label{sec:model}

Our model is based on two graphs, a logical one and a physical one. For simplicity, we describe it with reference to unidirectional traffic; notice however that our model and our results also account for bidirectional traffic. 
\Tab{notation} summarizes all the notation we introduce below.

\subsection{The logical graph}

The {\em logical} graph, exemplified in \Fig{logical}, describes where, i.e., which endpoint, the traffic comes from, and how it is processed. Its vertices are either {\em endpoints}~$e\in\Ec$ or {\em VNFs}~$v\in\Vc$. With reference to \Fig{logical}, we have~$\Ec=\{\text{RRH}\}$, and~$\Vc=\{\text{eNB},\text{P/S-GW},\text{MME},\text{HSS}\}$.

On the logical graph, we have logical flows~$l(e,v_1,v_2)$ representing data originating from endpoint~$e$ and going from VNF~$v_1$ to VNF~$v_2$. Additionally, with an abuse of notation, we indicate with~$l(e,v)$ flows that start from endpoint~$e$ and are first processed at VNF~$v$, e.g., from the RRH to the eNB in \Fig{logical}.
Note that keeping track of the endpoint at which flows originate, i.e., having an~$e$ index in our variables, serves a manifold purpose. First, it allows our model to account for the fact that different types of traffic (i.e., originating from different endpoints) may need different processing, i.e., traverse different VNF graphs. Furthermore, such VNF graphs may overlap; in this case, keeping track of the origin of the flows makes it possible to distinguish them even if they traverse the same VNF. Finally, it allows routing each flow in a different way, in both the logical and the physical graph. Notice that different traffic flows coming from the same physical endpoint can be distinguished by associating them to different {\em logical} endpoints.

Another important aspect of the system is that {\em there is no flow conservation in the logical graph}. As an example, in \Fig{logical} we see a user flow of $1$~traffic unit going from the RRH to eNB and thence to the gateway, which triggers some additional control traffic from the eNB and the gateway to the MME. Indeed, the following {\em generalized flow conservation} law holds for each endpoint~$e$ and VNFs~$v_2,v_3$:
\begin{equation}
l(e,v_2,v_3){=}\sum_{v_1\in\Vc}l(e,v_1,v_2)\chi(v_1,v_2,v_3)+l(e,v_2)\chi(e,v_2,v_3).\nonumber
\end{equation}
The above expression represents  the logical flow originated at endpoint $e$,  outgoing from VNF~$v_2$ and directed to VNF~$v_3$. Such a quantity is equal to the sum between logical flows entering~$v_2$, from either a VNF~$v_1$ or the endpoint~$e$ itself, multiplied by a factor~$\chi$. In particular, $\chi(v_1,v_2,v_3)$ is used to quantify the amount of logical flow directed to~$v_3$ that is generated when  traffic  coming from~$v_1$ is processed at VNF~$v_2$. With reference to the eNB in \Fig{logical}, we have~$\chi(\text{RRH},\text{eNB},\text{P/S-GW})=1$, while~$\chi(\text{RRH},\text{eNB},\text{MME})=0.3$. Similarly, for the gateway, we have~$\chi(\text{eNB},\text{P/S-GW},\text{MME})=0.2$. At the MME we have flow conservation, i.e.,~$\chi(\text{eNB},\text{MME},\text{HSS})=\chi(\text{P/S-GW},\text{MME},\text{HSS})=1$. In $\chi(e,v_2,v_3)$,  we abuse the notation and allow the first index of~$\chi$ to be an endpoint instead of a VNF. 
We remark that  $\chi$-values lower than one can also represent, e.g., a firewall dropping some of the incoming traffic. Also notice that $\chi$-values different from one can happen for both control traffic (e.g., the eNB in \Fig{logical}) and user traffic (as in the case of the firewall).

\begin{figure}[t]
\centering
\includegraphics[width=.9\columnwidth]{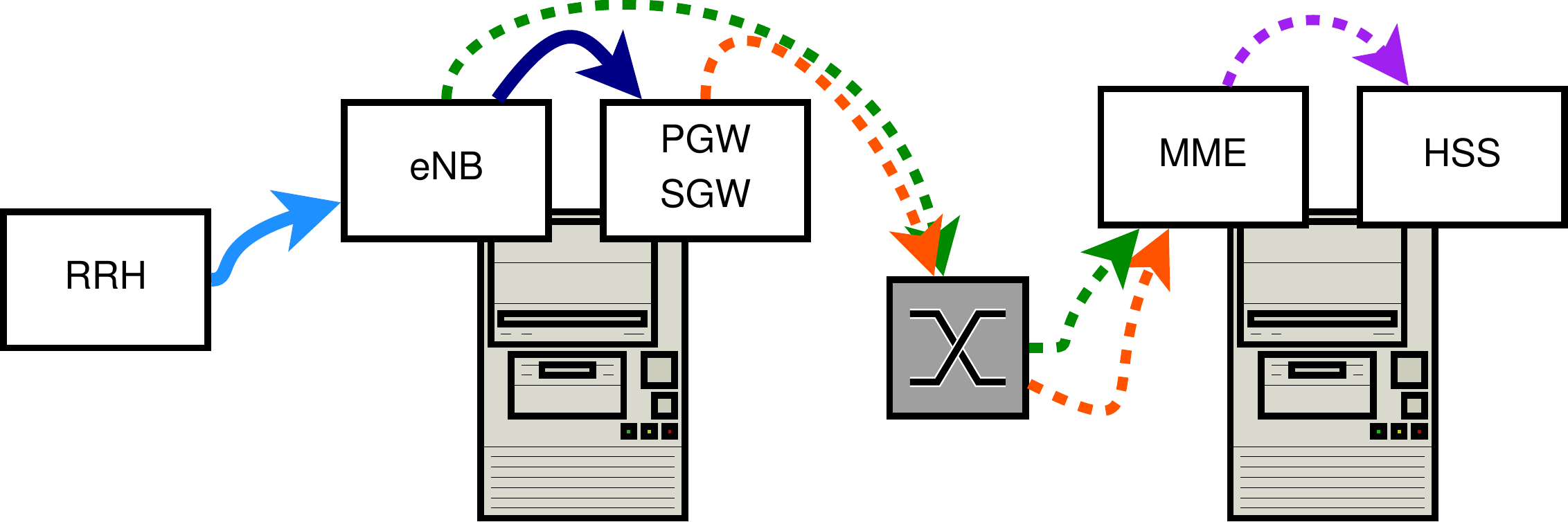}
\caption{
    Example implementation of the logical graph in \Fig{logical} over a physical network. Each line corresponds to a physical flow, i.e., to a $\tau$-variable; their color and style match the logical flows in \Fig{logical}.
    \label{fig:example}
} 
\end{figure}

\subsection{The physical graph}

In the physical graph, vertices correspond to the endpoints~$e\in\Ec$ and the B/F nodes~$c\in\Cc$. In general, B/F nodes  have computational capabilities~$k(c)$; B/F nodes that cannot host any VNF (e.g., switches) have~$k(c)=0$. \Fig{example} presents a possible implementation of the logical VNF graph in \Fig{logical}, where VNFs are placed on each of the two B/F nodes with processing capabilities.
For simplicity, we present our model with reference to the case where multiple VNF instances can be deployed across different nodes, but at most one instance of each VNF can be deployed at each B/F node.

Traffic traversing link~$(i,j)\in\Lc\subseteq(\Cc\cup\Ec)^2$ is also subject to a network delay~$D_{i,j}$. 
Such a delay is static, i.e., every unit of traffic traversing link~$(i,j)$ incurs a delay~$D_{i,j}$. Furthermore, links~$(i,j)$ have a bandwidth~$B_{i,j}$, corresponding to the maximum amount of traffic that can go from B/F node~$i$ to B/F node~$j$ without generating congestion.

Our main variable is represented by {\em physical flows}~$\tau_{i,j}(e,v_1,v_2)$, representing the amount of traffic that was originated from endpoint~$e$, last visited VNF~$v_1$, will next visit VNF~$v_2$, and is now traveling on link~$(i,j)$. 
Recall that we have to keep track of the flow originating endopint, in order to model traffic routing. 
If the flow has never been processed, i.e., it is going from~$e\in\Ec$ to its first VNF~$v\in\Vc$, we will conventionally set~$v_1=v_2=v$ and write~$\tau_{i,j}(e,v,v)$.

Given a B/F node~$c\in\Cc$, we denote by~$t_c(e,v_1,v_2)$ the amount of traffic that is just {\em transiting} by~$c$ (i.e., it is {\em not} processed at $c$) and it was originated at~$e$,  last visited VNF~$v_1$ and will next visit VNF~$v_2$. Similarly, $p_c(e,v_1,v_2)$ is the traffic that {\em is processed} at B/F node~$c$,  it was originated at~$e$, and  last visited VNF~$v_1$. Note that $p_c(e,v_1,v_2)>0$ implies that an instance of VNF $v_2$ is deployed at~$c$. 

Traffic being processed at VNF~$v$ is subject to a delay~$D(v)$. Normally,  processing delay is linked to the amount of resources (e.g., CPU) allocated to each VNF, and such an amount depends on the other VNFs deployed at the same B/F node. In our case, however, energy is the main metric of interest, and we can therefore assume that no VNF will be allocated more resources than  the minimum amount required by the VNF itself.

A first constraint we need to impose is that, given a generic VNF $v_2$, the traffic originated at $e$, that has been processed through VNF $v_1$ and is entering B/F node~$c$, is  either (i) processed at an instance of $v_2$ located in $c$, or (ii) transiting by $c$ while being routed toward an instance of $v_2$. Thus, for any $c, e, v_1,v_2$, we have:
\begin{equation}
\label{eq:flow-in}
\sum_{(i,c)\in\Lc}\tau_{i,c}(e,v_1,v_2)=t_c(e,v_1,v_2)+p_c(e,v_1,v_2).
\end{equation}
A similar constraint concerns the traffic outgoing from $c$. For any $c$, endpoint~$e$ and VNFs~$v_2,v_3$, we have:
\begin{equation}
\label{eq:flow-out}
\sum_{(c,j)\in\Lc}\hspace{-2mm}\tau_{c,j}(e,v_2,v_3){=}t_c(e,v_2,v_3)\mathord{+}\hspace{-2mm}\sum_{v_1\in\Vc}\hspace{-2mm}p_c(e,v_1,v_2)\chi(v_1,v_2,v_3)
\end{equation}
where $v_2$ is the last VNF that traffic visited, either before arriving at~$c$ (if traffic just transits by $c$) or at~$c$ itself (if $v_2$ is deployed therein, i.e., $p_c(e,v_1,v_2)>0$). $v_3$ instead is the VNF that  traffic will visit next.
In other words, \Eq{flow-in}--\Eq{flow-out} enforce {\em ordinary} flow conservation for the traffic that is transiting at~$c$, i.e., using $c$ as a traditional switch, and {\em generalized} flow conservation for the traffic that is processed at~$c$. 

Next, we need to ensure that we only use active B/F nodes and links, and their capacity is not exceeded. We define two sets of binary variables,~$x_{i,j}$ and~$y_c$, indicating whether link~$(i,j)$ and B/F node~$c$ are active or not.

For links, we need to impose: 
\begin{equation}
\label{eq:enable-link}
x_{i,j}\leq \min\left\{y_i,y_j\right\},\quad\forall (i,j)\in\Lc\,,
\end{equation}
i.e., no link can be active if either of its ends is off, and
\begin{equation}
\label{eq:capacity-l}
\sum_{e\in\Ec}\sum_{v_1,v_2\in\Vc}\tau_{i,j}(e,v_1,v_2)\leq x_{i,j}B_{i,j},\quad\forall (i,j)\in\Lc.
\end{equation}

\begin{table*}[t]
\caption{Notation}
\label{tab:notation} 
\centering
\begin{tabularx}{1\textwidth}{|p{1.5cm}|p{2cm}|X|}
\hline
Symbol & Type & Meaning\\
\hline\hline
$\Ec$ & Set & Set of network endpoints\\
\hline
$\Cc$ & Set & Set of B/F nodes\\
\hline
$\Lc$ & Set & Set of links\\
\hline
$\Vc$ & Set & Set of VNFs\\
\hline
$B_{i,j}$ & Parameter & Bandwidth of link~$(i,j)\in\Lc$\\
\hline
$D_{i,j}$ & Parameter & Delay of link~$(i,j)\in\Lc$\\
\hline
$\chi(v_1,v_2,v_3)$ & Parameter & How much traffic resulting from the processing at VNF~$v_2$,  which  was previously processed at VNF~$v_1$,  is meant to be next processed at VNF~$v_3$\\
\hline
$\delta(c,v)$ & Binary var. & Whether we deploy VNF~$v\in\Vc$ at B/F node~$c\in\Cc$\\
\hline
$f_0$ & Function & Energy consumption due to placing a VNF at a B/F node\\
\hline
$f_\text{idle}$ & Function & Energy consumption due to activating a B/F node\\
\hline
$f_\text{proc}$ & Function & Traffic-dependent energy consumption due to processing\\
\hline
$f_\text{sw}$, $f_\text{link}$ & Function & Traffic-dependent energy consumption at switches and links\\
\hline
$k(c)$ & Parameter & Computational capability of B/F node~$c\in\Cc$\\
\hline
$l(e,v_1,v_2)$ & Parameter & Logical flow originated at~$e\in\Ec$ and going from VNF~$v_1\in\Vc$ to VNF~$v_2\in\Vc$\\
\hline
$l(e,v)$ & Parameter & Logical flow originating at~$e\in\Ec$ and first being processed at VNF~$v\in\Vc$\\
\hline
$p_c(e,v_1,v_2)$ & Continuous var. & How much traffic coming from users connected to endpoint~$e\in\Ec$ for service that was last processed at VNF~$v_1$ is processed by an instance of VNF~$v_2$ deployed at B/F node~$c$\\
\hline
$r(v)$ & Parameter & Computational capability required to process one traffic unit of VNF~$v\in\Vc$\\
\hline
$\rho(c)$ & Parameter & Computational capability consumed by one unit of traffic transiting by B/F node (SW switch) $c\in\Cc$\\
\hline$\tau_{i,j}(e,v_1,v_2)$ & Continuous var. & How much traffic coming from users connected to endpoint~$e\in\Ec$ that was last processed at VNF~$v_1$ and meant to be next processed at VNF~$v_2$ goes through link~$(i,j)\in\Lc$\\
\hline
$t_c(e,v_1,v_2)$ & Continuous var. & How much traffic originating from~$e$ that was last processed at VNF~$v_1$ and meant to be next processed at VNF~$v_2$ transits (without processing) by B/F node~$c\in\Cc$\\
\hline
$x_{i,j}$ & Binary var. & Whether link~$(i,j)\in\Lc$ is active\\
\hline
$y_c$ & Binary var. & Whether B/F node~$c\in\Cc$ is active\\
\hline
\end{tabularx}
\end{table*}

With regard to processing, 
inactive B/F nodes cannot host any VNF. We track this through a binary variable~$\delta(c,v)$ expressing whether an instance of VNF~$v$ is deployed at B/F node~$c$, and impose:
\begin{equation}
\label{eq:enable-core}
\delta(c,v)\leq y_c,\quad\forall c\in\Cc,v\in\Vc.
\end{equation}
Additionally, no processing can be done for VNFs that are not deployed at a given B/F node: 
\begin{equation}
\label{eq:honor-delta}
p_c(e,v_1,v_2)\leq \delta(c,v_2)k(c),\quad\forall c\in\Cc,e\in\Ec,v_1, v_2\in\Vc.
\end{equation}
Finally,  {\em each traffic unit} processed by VNF~$v$ requires~$r(v)$ computational capability, and, assuming~$c$ is a software switch, each unit of traffic switched by~$c$ consumes~$\rho(c)$ CPU.  
Clearly, the computational capability of each B/F node~$c$ must be sufficient for both, i.e., for any B/F node~$c$,
\begin{multline}
\sum_{e\in\Ec}\sum_{v_1\in\Vc}\sum_{v_2\in\Vc}\Bigg[r(v_2)p_c(e,v_1,v_2)+\\
 +\rho(c)\sum_{(c,j)\in \Lc}\tau_{c,j}(e,v_1,v_2)\Bigg]\leq k(c),
\label{eq:capacity-c}
\end{multline}
where $\rho(c)$ multiplies the total traffic outgoing from $c$.

Next, we ensure that the delay of the traffic originated at any endpoint~$e$ does not exceed a threshold~$D^{\max}(e)$:
\begin{multline}
\label{eq:delay}
\frac{\sum_{i,j\in\Lc}\sum_{v_1,v_2\in\Vc}D_{i,j}\tau_{i,j}(e,v_1,v_2)}{\sum_{v\in\Vc}l(e,v)}+\\
+\frac{\sum_{v_1,v_2\in\Vc}\sum_{c\in\Cc}D(v_2)p_c(e,v_1,v_2)}{\sum_{v\in\Vc}l(e,v)}\leq D^{\max}(e).
\end{multline}
The two terms on the left hand side of \Eq{delay} correspond to the network and processing delay, respectively. 
The first term of \Eq{delay} is a summation of terms in the form~$D_{i,j}\frac{\tau_{i,j}}{l}$, each representing the delay incurred by traversing link~$(i,j)$ weighted by the fraction of traffic traversing it. Similarly, the second term of \Eq{delay} is a summation of terms in the form~$D(v)\frac{p_c}{l(e,v)}$, weighting the processing delay of VNF~$v$ by the fraction of traffic processed by it.

At last, logical and physical flows have to match. To this end, it is sufficient to impose that, for each logical flow~$l(e,v)$ going from endpoint~$e$ to VNF~$v$, there are corresponding physical flows of the type~$\tau_{e,j}(e,v,v)$, such that:
\begin{equation}
\label{eq:match-1}
l(e,v)=\sum_{(e,j)\in\Lc}\tau_{e,j}(e,v,v),\quad\forall e\in\Ec,v\in\Vc.
\end{equation}
Eq.\,\Eq{match-1} ensures that the traffic injected from endpoints to B/F nodes on the physical graph matches the logical traffic going from endpoints to VNFs. Thanks to the flow conservation constraints \Eq{flow-in}--\Eq{flow-out}, this also implies that such traffic is processed and transformed as dictated by the $\chi$-parameters, i.e., that all physical flows match their logical counterpart.

\subsection{Energy and objective}

There are five contributions to the overall energy consumption we are interested in tracking:
\begin{itemize}
    \item activating a B/F node, resulting in a consumption of~$f_\text{idle}$;
    \item placing a VNF on a B/F node, resulting in a consumption~$f_0$ due to, e.g., virtual machines overhead;
    \item using said VNF, resulting in a consumption of~$f_\text{proc}$ depending on the computational resources used;
\item switching traffic at a B/F node, resulting in a consumption of~$f_\text{sw}$ depending on the traffic switched by the node; 
    \item having traffic going through links, resulting in a consumption of~$f_\text{link}$ depending on the traffic over each link.
\end{itemize}
The corresponding energy consumption is:
\begin{equation}
E_\text{idle}=\sum_{c\in\Cc}f_\text{idle}\left(y_c\right); \quad \quad E_0=\sum_{c\in\Cc}\sum_{v_2\in\Vc}f_0\left(\delta(c,v_2)\right); \nonumber
\end{equation}
\begin{equation}
E_\text{proc}=\sum_{c\in\Cc}f_\text{proc}\left ( \sum_{v_2\in\Vc}r(v)\sum_{e\in\Ec}\sum_{v_1\in\Vc}p_c(e,v_1,v_2)\right ); \nonumber
\end{equation}
\begin{equation}
E_\text{sw}=\sum_{c\in\Cc}f_\text{sw}\left (\sum_{e\in\Ec}\sum_{v_1,v_2\in\Vc} \tau_{c,j}(e,v_1,v_2) \right ); \nonumber
\end{equation}
\begin{equation}
E_\text{link}=\sum_{(i,j)\in\Lc}f_\text{link}\left (\sum_{e\in\Ec}\sum_{v_1,v_2\in\Vc}\tau_{i,j}(e,v_1,v_2) \right ). \nonumber
\end{equation}
Given all this, our objective can be written as:
\begin{equation}
\label{eq:obj}
\min_{x,y}E=E_0+E_\text{proc}+E_\text{idle}+E_\text{sw}+ E_\text{link},
\end{equation} 
subject to the following constraints:
\footnotesize{
\begin{equation}
\nonumber
l(e,v_2,v_3){=}\sum_{v_1\in\Vc}l(e,v_1,v_2)\chi(v_1,v_2,v_3)+l(e,v_2)\chi(e,v_2,v_3)
\end{equation}
\begin{equation}
\nonumber
\sum_{(i,c)\in\Lc}\tau_{i,c}(e,v_1,v_2)=t_c(e,v_1,v_2)+p_c(e,v_1,v_2)
\end{equation}
\begin{equation}
\nonumber
\sum_{(c,j)\in\Lc}\hspace{-2mm}\tau_{c,j}(e,v_2,v_3){=}t_c(e,v_2,v_3)\mathord{+}\hspace{-2mm}\sum_{v_1\in\Vc}\hspace{-2mm}p_c(e,v_1,v_2)\chi(v_1,v_2,v_3)
\end{equation}
\begin{equation}
\nonumber
\sum_{e\in\Ec}\sum_{v_1,v_2\in\Vc}\tau_{i,j}(e,v_1,v_2)\leq x_{i,j}B_{i,j}
\end{equation}
\begin{equation}
\nonumber
\delta(c,v)\leq y_c\quad;\quad l(e,v)=\sum_{(e,j)\in\Lc}\tau_{e,j}(e,v,v)
\end{equation}
\begin{equation}
\nonumber
p_c(e,v_1,v_2)\leq \delta(c,v_2)k(c) \quad ; \quad x_{i,j}\leq \min\left\{y_i,y_j\right\}
\end{equation}
\begin{equation}
\nonumber
\sum_{e\in\Ec}\sum_{v_1\in\Vc}\sum_{v_2\in\Vc}\left[r(v_2)p_c(e,v_1,v_2){+}
\rho(c)\sum_{(c,j)\in \Lc}\tau_{c,j}(e,v_1,v_2)\right]{\leq} k(c)
 \end{equation}
\begin{multline}
\nonumber
 \frac{\sum_{i,j\in\Lc}\sum_{v_1,v_2\in\Vc}D_{i,j}\tau_{i,j}(e,v_1,v_2)}{\sum_{v\in\Vc}l(e,v)}+ \\
+\frac{\sum_{v_1,v_2\in\Vc}\sum_{c\in\Cc}D(v_2)p_c(e,v_1,v_2)}{\sum_{v\in\Vc}l(e,v)}\leq D^{\max}(e)
\end{multline}
} 

\normalsize

\subsection{Multiple VNF instances}

So far, we have presented our problem formulation with reference to the case that at most one instance of each VNF can be placed at each B/F node. This is true in many cases; however, there are situations (e.g., coexisting services with isolation requirements) when we may need to place multiple instances of the same VNF at the same B/F node. In the following, we extend our model to describe such a case.

To begin with, we need to distinguish VNFs from VNF instances. To this end, we introduce a new set~$\Wc=\{w\}$ representing the VNF instances, and indicate as~$V(w)\in\Vc$ the type of instance~$w$, i.e., the VNF~$w$ is an instance of.

Furthermore, we need to account for the fact that logical flows happen between VNFs, while physical flows happen between VNF instances and processing takes place at VNF instances. Therefore, we need to replace:
\begin{itemize}
    \item $\tau_{i,c}(e,v_1,v_2)$ with $\tau_{i,c}(e,w_1,w_2)$, where~$w_1,w_2\in\Wc$;
    \item $t_c(e,v_1,v_2)$ with $=t_c(e,w_1,w_2)$;
    \item $p_c(e,v_1,v_2)$ with $=p_c(e,w_1,w_2)$.
\end{itemize}

In order to guarantee that physical and logical flows match, we also need to replace \Eq{match-1} with:
\begin{equation}
\label{eq:match-1-multi}
l(e,v)=\sum_{(e,j)\in\Lc}\sum_{w\in\Wc:V(w)=v}\tau_{e,j}(e,w,w),\quad\forall e\in\Ec,v\in\Vc,
\end{equation}
where, in \Eq{match-1-multi}, the second summation accounts for all instances~$w$ of VNF~$v$.

Finally, \Eq{flow-out} needs to be changed in order to represent the fact that that data can flow from any instance of a VNF to any instance of the next VNF in the logical graph:
\begin{multline}
\label{eq:flow-out-multi}
\sum_{(c,j)\in\Lc}\sum_{\substack{w_2,w_3\in\Wc\colon\\V(w_2)=v_2\\V(w_3)=v_3}}\tau_{c,j}(e,w_2,w_3)=\sum_{\substack{w_2,w_3\in\Wc\colon\\V(w_2)=v_2\\V(w_3)=v_3}}t_c(e,w_2,w_3)+\\
+\sum_{\substack{w_1\in\Wc\colon\\V(w_1)=v_1}}\sum_{\substack{w_2,w_3\in\Wc\colon\\V(w_2)=v_2\\V(w_3)=v_3}}p_c(e,w_1,w_2)\chi(v_1,v_2,v_3).
\end{multline}
In \Eq{flow-out-multi}, notice how the $\chi$-variable, which concerns logical flows, has as its indices VNFs in~$\Vc$, while the $\tau$- and $p$-variables have as indices VNF instances in~$\Wc$.

\section{The OptiLoop strategy}
\label{sec:heur}

The problem stated in \Sec{model} falls into the MILP category, and is thus impractical to solve in real time. We can however solve its  {\em relaxed} version, where binary variables are allowed to take any value in~$[0,1]$. Optimal solutions to the relaxed models cannot be directly used to manage (or plan) a network; however, they can provide useful guidelines.

\begin{figure}
\centering
\includegraphics[width=1\columnwidth]{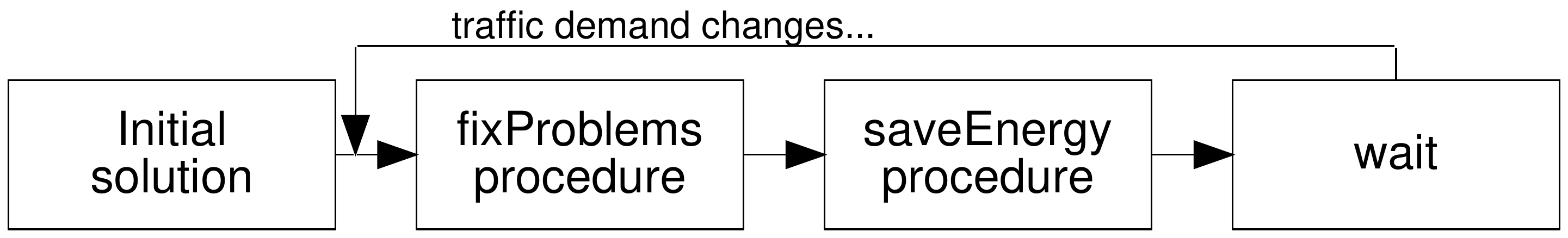}
\caption{
    The OptiLoop strategy. We begin by obtaining an initial feasible solution, as described in \Sec{initial}. After that, we periodically check the current solution for problems (procedure \protect\path{fixProblems}, described in \Alg{fixproblems}) and for opportunities to deactivate some B/F nodes and/or links (procedure \protect\path{saveEnergy}, described in \Alg{saveenergy}).
    \label{fig:online}
} 
\end{figure}

Our basic idea of is to leverage the software-defined nature of our network to make an optimizer {\em interact} with SDN controllers and NFVOs, i.e., optimize problems as a part of our network management strategy. Our solution strategy is called OptiLoop (for Optimization in the Loop) and it includes the following steps, as outlined in \Fig{online}: (i) we initialize the system with a feasible (albeit potentially suboptimal) solution, as detailed in \Sec{initial}; (ii) after that, we periodically:
    \begin{enumerate}
        \item check that the network configuration is adequate to the current (and/or predicted) demand;
        \item if not so, activate additional VNFs, B/F nodes, and/or links as needed;
        \item check whether there are B/F nodes and/or links that can be deactivated in order to save energy;
        \item if so, update the current network configuration accordingly.
    \end{enumerate}
\Sec{initial} explains how we obtain the initial solution, i.e., Item~(i) above. Items (1)--(2) and (3)--(4) correspond to the \path{fixProblems} and \path{saveEnergy} procedures respectively, which are described in \Sec{fixproblems} and \Sec{saveenergy}.

It is worth pointing out that the \path{fixProblems} and \path{saveEnergy} procedures are designed to take no action if no action is warranted, and therefore there is no harm in cascading them. As an example, \path{fixProblems} will never take any action the first time it is executed after an initial solution is generated, as that solution is guaranteed to be feasible. Similarly, \path{saveEnergy} is unlikely to find elements to deactivate if \path{fixProblems} just had to activate some.

\subsection{Initial solution}
\label{sec:initial}

The initial solution used to initialize OptiLoop  has to be feasible, but does not have to be optimal. It can come from one of the heuristics we reviewed in \Sec{relwork}, or it can be obtained by solving a version of our problem where:
\begin{enumerate}
    \item all B/F nodes and links are active, i.e.,~$y_{c}=1,\forall c\in\Cc$ and~$x_{i,j}=1,\forall (i,j)\in\Lc$;
    \item there is an instance of all VNFs deployed at each B/F node, i.e.,~$\delta(c,v)=1,\forall c\in\Cc,v\in\Vc$.
\end{enumerate}

The resulting solution will be highly suboptimal, as we are likely to needlessly activate B/F nodes and/or links and to place useless VNF instances, all of which increase the power consumption. On the plus side, the problem is LP, as all binary variables are fixed; furthermore, the following property holds.
\begin{property}
\label{prop:allone}
If a problem instance is feasible, then there is at least one feasible solution where the $x,y$ and~$\delta$ variables are all set to~$1$.
\end{property}
\begin{IEEEproof}
Let us consider a feasible solution~$\Sc^0$, where some of the binary variables are set to zero and others to one. By hypothesis,~$\Sc^0$ is feasible. What we need to prove is that changing {\em all} binary variables to one can never make us violate a constraint. This follows by inspection of \Eq{enable-link}, \Eq{capacity-l}, \Eq{enable-core}, \Eq{honor-delta}: if they hold for the variable values in~$\Sc^0$, then they will also hold when all binary variables are set to one.
\end{IEEEproof}
In other words, setting all binary variables to one is an easy way to obtain a feasible solution to our problem to start with. This  solution can be vastly improved, as discussed next.

\subsection{The \protect\path{fixProblems} procedure}
\label{sec:fixproblems}

The high-level goal of the \path{fixProblems} procedure is to check whether the current network configuration can cope with the current (and projected) traffic demand. If this is not the case, then we take one or more of the following actions: (i) activating additional B/F nodes; (ii)  activating additional links; (iii) deploying additional VNF instances.

Specifically, as detailed in \Alg{fixproblems}, we take as an input the current solution~$\Sc^\text{curr}$.
We then proceed, in \Line{new}--\Line{fix0-d}, to create a new instance~$\Pc$ of the problem, where all binary variables are fixed to their values in~$\Sc^\text{curr}$. In \Line{opt0}, we solve such a problem: if it is feasible, then no action is required and the algorithm exits (\Line{ok-exit}). Otherwise, we look at why the problem is infeasible, by inspecting its {\em irreducible inconsistent subsystem} (IIS), i.e., the subset of constraints such that removing any of them would make the problem feasible. This set allows us to discriminate between the different reasons that can make the network unable to operate properly (hence, the problem infeasible).

\begin{algorithm}[t]
\caption{
The \protect\path{fixProblems} procedure.
    \label{alg:fixproblems}
}
\begin{algorithmic}[1]
\Require{$\Sc^\text{curr}$} \label{line:input}

\State{$\Pc\gets\textbf{new}\text{ problem}()$} \label{line:new}

\State{$\Pc.\text{fix}(x_{i,j}\gets x^\text{curr}_{i,j},\quad\forall (i,j)\in\Lc)$} \label{line:fix0-a}
\State{$\Pc.\text{fix}(y_c\gets y^\text{curr}_c,\quad\forall c\in\Cc)$}
\State{$\Pc.\text{fix}(\delta(c,v)\gets\delta^\text{curr}(c,v),\quad\forall c\in\Cc,v\in\Vc)$} \label{line:fix0-d}
\State{$\textbf{solve}(\Pc)$} \label{line:opt0}
\If{$\Pc.\text{is\_feasible}$}
 \State\Return \label{line:ok-exit}
\EndIf

\If{\Eq{capacity-l}$\in\Pc.\text{IIS}$}
 \State{$\Pc.\text{relax}(x_{i,j}\colon x^\text{curr}_{i,j}=0)$} \label{line:probl-rx}
 \State{$\Pc.\text{relax}(y_c\colon y^\text{curr}_c=0)$} \label{line:probl-ry}
 \State{$\tilde{x},\tilde{y}\gets\textbf{solve}(\Pc)$} \label{line:probl-solve}
 \State{$(i^\star,j^\star){\gets}\textbf{choose}\text{ from }\Lc\text{ with prob. }\tilde{x}_{i,j}$} \label{line:probl-choose}
 \State{$\Pc.\text{fix}(x_{i^\star,j^\star}\gets 1)$} \label{line:probl-fx}
 \State{$\Pc.\text{fix}(y_i\gets 1;y_j\gets 1)$} \label{line:probl-fy}
 \State{{\bf goto} \Line{opt0}} \label{line:probl-goto}
\EndIf

\If{\Eq{capacity-c}$\in\Pc.\text{IIS}$}
 \State{$\Pc.\text{relax}(y(c)\colon y^\text{curr}(c)=0)$} \label{line:probs-ry}
 \State{$\Pc.\text{relax}(\delta(c,v)\colon \delta^\text{curr}(c,v)=0)$} \label{line:probs-rd}
 \State{$\tilde{\delta}\gets\textbf{solve}(\Pc)$} \label{line:probs-solve}
 \State{$c^\star,v^\star{\gets}\textbf{choose}\text{ from }\Cc\times\Vc\text{ with prob. }\tilde{\delta}(c,v)$} \label{line:probs-choose}
 \State{$\Pc.\text{fix}(y(c^\star)\gets 1)$} \label{line:probs-fixy}
 \State{$\Pc.\text{fix}(\delta(c^\star,v^\star)\gets 1)$} \label{line:probs-fixd}
 \State{{\bf goto} \Line{opt0}} \label{line:probs-goto}
\EndIf

\end{algorithmic}
\end{algorithm}

If constraint \Eq{capacity-l} (mandating that no link is used for more than its capacity) is in the IIS, then we need to activate some more links and/or B/F nodes. To decide which ones, we relax all~$x$- and $y$-variables related to B/F nodes and links that were inactive in~$\Sc^\text{curr}$ (\Line{probl-rx}--\Line{probl-ry}) and solve the new problem (\Line{probl-solve}). We then choose one link to activate, with a probability proportional to its relaxed~$\tilde{x}_{i,j}$ value, and fix to~1 the corresponding $x$-value 
and the $y$-values of its endpoints (\Line{probl-choose}--\Line{probl-fy}). We then go back to \Line{opt0} and test the new solution (\Line{probl-goto}). If it is still infeasible, we will activate further network elements until feasibility is achieved.

We proceed in a similar way if constraint \Eq{capacity-c} is in the IIS, i.e., if we have  a computational capability issue. We relax variables~$y$ and~$\delta$, allowing for more B/F nodes to be activated and VNFs to be deployed if needed, and solve the new problem obtaining the relaxed values~$\tilde{\delta}$ (\Line{probs-ry}--\Line{probs-solve}). We then have to decide which VNF to place and where. We do so by selecting a B/F node~$c^\star$ and a VNF~$v^\star$ at random, with a probability proportional to the relaxed values~$\tilde{\delta}(c,v)$, and fix the corresponding~$y$ and $\delta$-variable to~$1$ 
(\Line{probs-choose}--\Line{probs-fixd}). Finally, we go back to testing the new solution (\Line{probs-goto}).

Note that all  problems we solve in \Alg{fixproblems} are LP: in  \Line{opt0}, \Line{probl-solve} and \Line{probs-solve} all binary variables are either fixed or relaxed. Such problems can be therefore solved in polynomial time ({\em embedded}~\cite{embedded} optimization on low-power hardware is now commonplace in several application domains).

\subsection{The \protect\path{saveEnergy} procedure}
\label{sec:saveenergy}

We can think of the \path{saveEnergy} procedure as the dual of \path{fixProblems}. Our aim is to identify B/F nodes and/or links that  can be deactivated, as well as VNF instances that can be removed from the B/F nodes they run into. The objective is to reduce our power consumption without impairing our ability to serve the traffic, i.e., without making the problem infeasible. As in the \path{fixProblems} procedure, we solve a sequence of LP problems with fixed or relaxed variables, obtaining guidance on the decisions we should make and their effects.

In \Alg{saveenergy}, we take the current solution~$\Sc^\text{curr}$ as an input. We then create an instance~$\Pc$ of the problem where the binary variables that in the current solution have value 0 are fixed to 0 (\Line{fixx2}--\Line{fixdelta2}), and those that have currently value 1 are relaxed (\Line{relaxx2}--\Line{relaxdelta2}). This is because we are not looking for new nodes/links to activate, but for elements to deactivate. We do so by solving the problem instance~$\Pc$ (\Line{solve2}); note that all binary variables therein are fixed or relaxed, so the problem is LP.

\begin{algorithm}[t]
\caption{
The \protect\path{saveEnergy} procedure.
    \label{alg:saveenergy}
}
\begin{algorithmic}[1]
\Require{$\Sc^\text{curr}$} \label{line:input2}

\State{$\Pc\gets\textbf{new}\text{ problem}()$} \label{line:new2}

\State{$\Pc.\text{fix}(x_{i,j}\gets 0,\quad\forall (i,j)\in\Lc\colon x^\text{curr}_{i,j}=0)$} \label{line:fixx2}
\State{$\Pc.\text{fix}(y_c\gets 0,\quad\forall c\in\Cc\colon y^\text{curr}_c=0)$}
\State{$\Pc.\text{fix}(\delta(c,v)\gets 0,\quad\forall c\in\Cc,v\in\Vc\colon\delta(c,v)=0)$} \label{line:fixdelta2}

\State{$\Pc.\text{relax}(x_{i,j},\quad\forall (i,j)\in\Lc\colon x^\text{curr}_{i,j}=1)$} \label{line:relaxx2}
\State{$\Pc.\text{relax}(y_c,\quad\forall c\in\Cc\colon y^\text{curr}_c=1)$}
\State{$\Pc.\text{relax}(\delta(c,v),\quad\forall c\in\Cc,v\in\Vc\colon\delta(c,v)=1)$} \label{line:relaxdelta2}

\State{$\textbf{solve}(\Pc)$} \label{line:solve2}

\State{$(x^\star,y^\star)\gets\arg\min_{(x,y)\in\Lc\colon x_{x,y}^\text{curr}=1}\tilde{x}_{i,j}$} \label{line:ijstar2}
\State{$c^\star\gets\arg\min_{c\in\Cc\colon y^\text{curr}(c)=1}\tilde{y}(c)$} \label{line:cstar2}
\State{$d^\star,v^\star\gets\arg\min_{c,v\in\Cc\times\Vc\colon \delta^\text{curr}(c,v)=1}\tilde{\delta}(c,v)$} \label{line:vstar2}

\State{$\Pc_2\gets\textbf{copy}(\Pc)$} \label{line:copy2}

\If{$\tilde{x}_{i^\star,j^\star}<\tilde{y}(c^\star) \wedge \tilde{x}_{i^\star,j^\star}<\tilde{\delta}(d^\star,v^\star)$} \label{line:if2-start}
 \State{$\Pc_2.\text{fix}(x_{i^\star,j^\star}\gets 0)$}
\EndIf

\If{$\tilde{y}(c^\star)<\tilde{x}_{i^\star,j^\star} \wedge \tilde{y}(c^\star)<\tilde{\delta}(d^\star,v^\star)$}
 \State{$\Pc_2.\text{fix}(y(c^\star)\gets 0)$}
 \State{$\Pc_2.\text{fix}(x_{i,j}\gets 0,\quad\forall (i,j)\in\Lc\colon i=c^\star\vee j=c^\star$)} \label{line:offx2}
 \State{$\Pc_2.\text{fix}(\delta(c,v)\gets 0,\quad\forall c\in\Cc,v\in\Vc\colon c=c^\star$)} \label{line:offd2}
\EndIf

\If{$\tilde{\delta}(d^\star,v^\star)<\tilde{x}_{i^\star,j^\star} \wedge \tilde{\delta}(d^\star,v^\star)<\tilde{y}(c^\star)$}
 \State{$\Pc_2.\text{fix}(\delta(d^\star,v^\star)\gets 0)$}
\EndIf \label{line:if2-end}

\State{$\textbf{solve}(\Pc_2)$} \label{line:solve22}

\If{$\Pc_2.\text{is\_feasible}$}
 \State{$\Pc\gets\Pc_2$} \label{line:pcgetp2}
 \State{{\bf goto} \Line{new2}} \label{line:goto2}
\Else
 \State\Return{$\Pc$}\label{line:returnp2}
\EndIf

\end{algorithmic}
\end{algorithm}

In \Line{ijstar2}--\Line{vstar2} we identify the link, B/F node, and pair of B/F node and VNF that are active in the current solution and have the lowest value of the associated relaxed variable (respectively~$\tilde{x}_{i,j}$, $\tilde{y}(c)$, and~$\tilde{\delta}(c,v)$). Intuitively, these are the  elements that most likely  can be  deactivated without impairing network functionality. We check this by creating a copy of problem instance~$\Pc$ and fixing to~$0$ the binary variable associated to the  element with the lowest value of the relaxed variables (\Line{copy2}--\Line{if2-end}).
If that element is a B/F node, we also need to deactivate the links using it and the VNF instances it hosts (\Line{offx2}--\Line{offd2}).

The difference between $\Pc$ and~$\Pc_2$ is that exactly one  element that was active in~$\Pc$ is deactivated in~$\Pc_2$, hence $\Pc_2$ is also LP. In \Line{solve22}, we solve~$\Pc_2$ and check if it is feasible. If that is the case, then we use~$\Pc_2$ as our new solution, and try to further enhance it (\Line{pcgetp2}--\Line{goto2}). Otherwise, the algorithm returns~$\Pc$, the last feasible solution we tried. 

In summary, \Alg{saveenergy} deactivates zero or more  elements, i.e., B/F nodes, links, or VNF instances. The element to deactivate is chosen based on the value taken by the corresponding relaxed variable, and after each change we check that the resulting  configuration  can serve its load, i.e., the problem instance is feasible.

\begin{figure*}
\centering
\subfigure[\label{fig:real-archi}]{
    \includegraphics[width=.7\columnwidth]{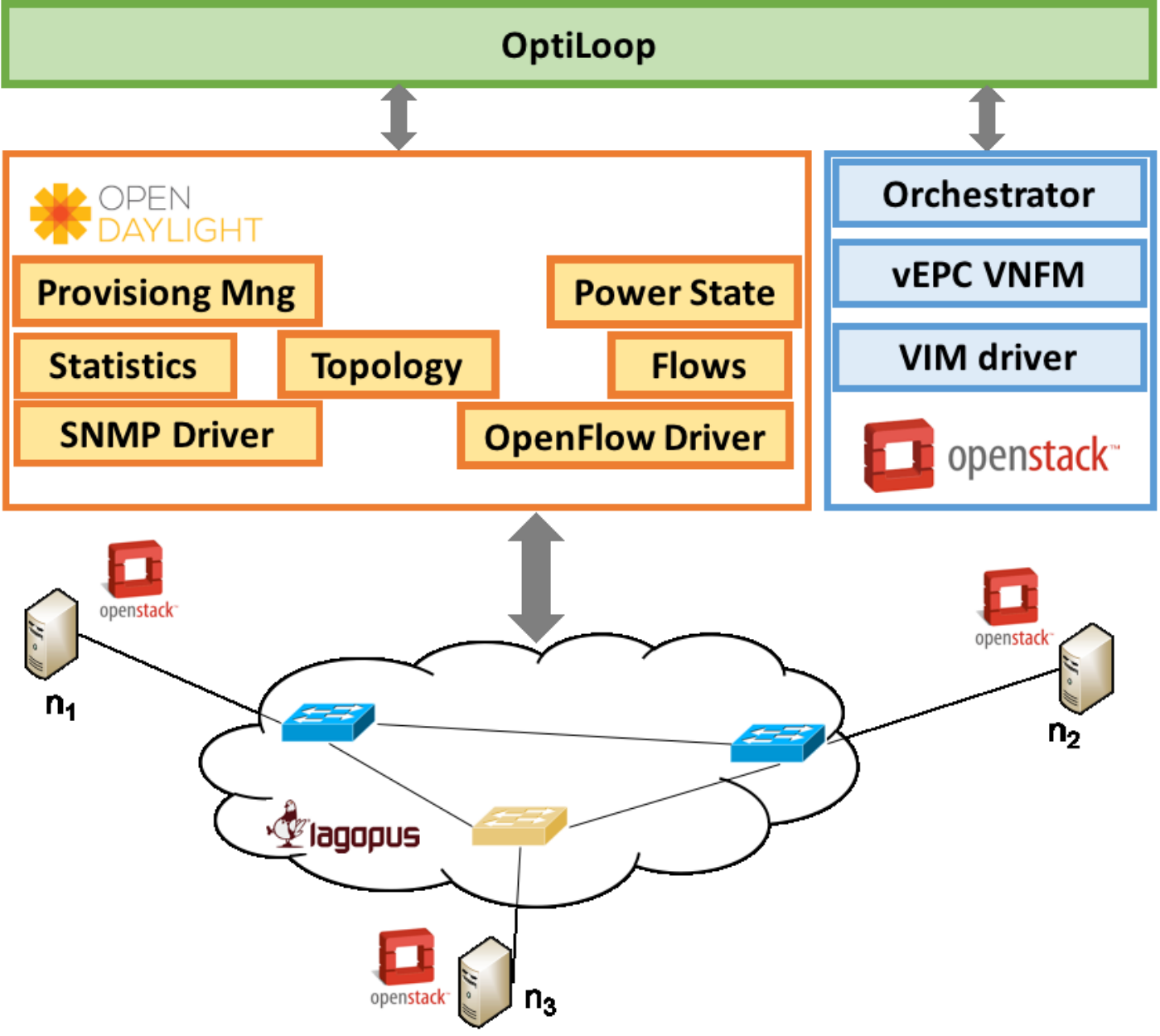}
} 
\hspace{1cm}
\subfigure[\label{fig:real-topo}]{
    \includegraphics[width=1\columnwidth]{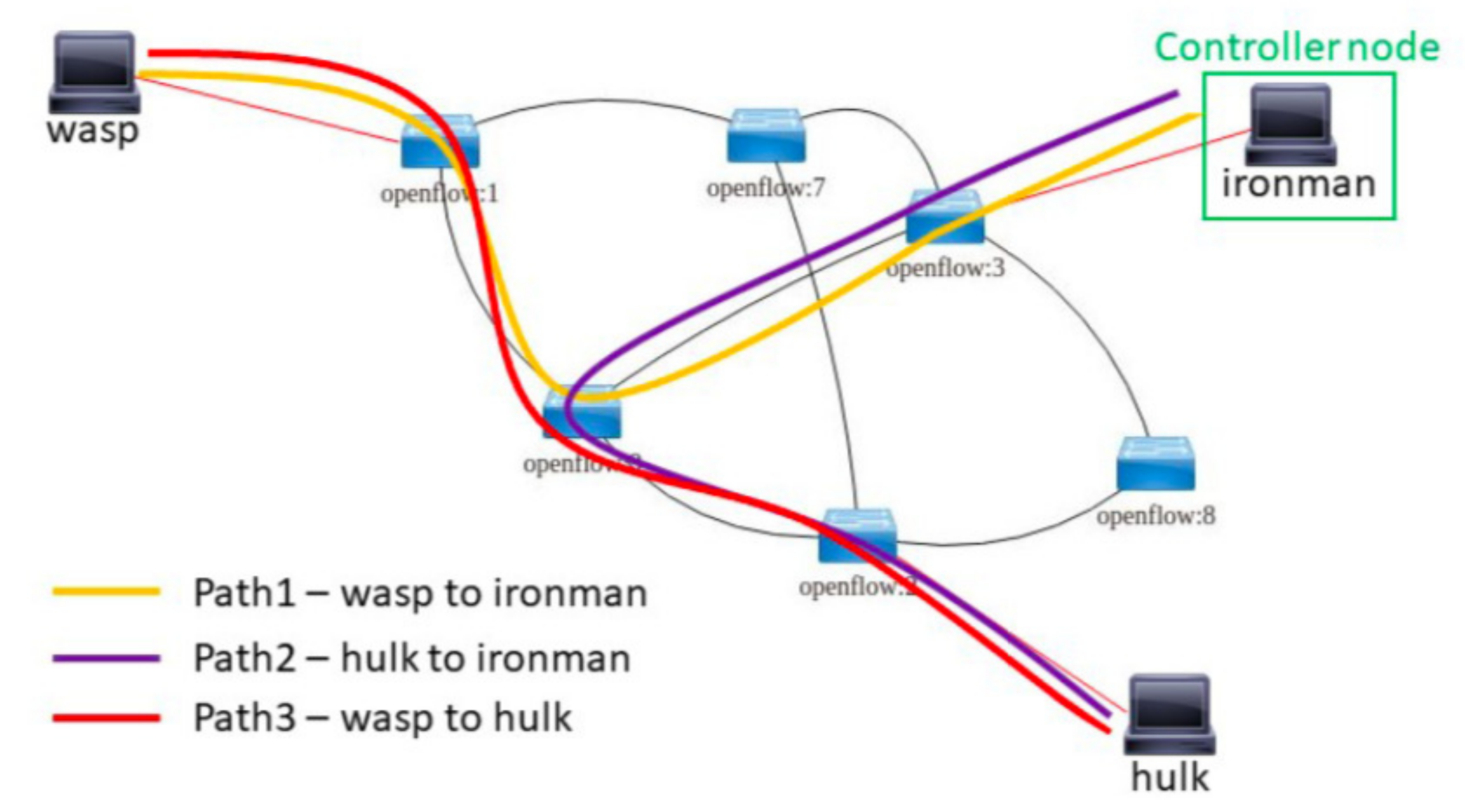}
} 
\caption{
    Architecture (left) and topology (right) of the real-world testbed.
    \Fig{real-topo} also indicates the paths used in our path instantiation experiments, discussed in \Sec{exp-path}.
    \label{fig:realtestbed}
} 
\end{figure*}

\begin{figure*}
\centering
\subfigure[\label{fig:mininet-archi}]{
    \includegraphics[width=.5\columnwidth]{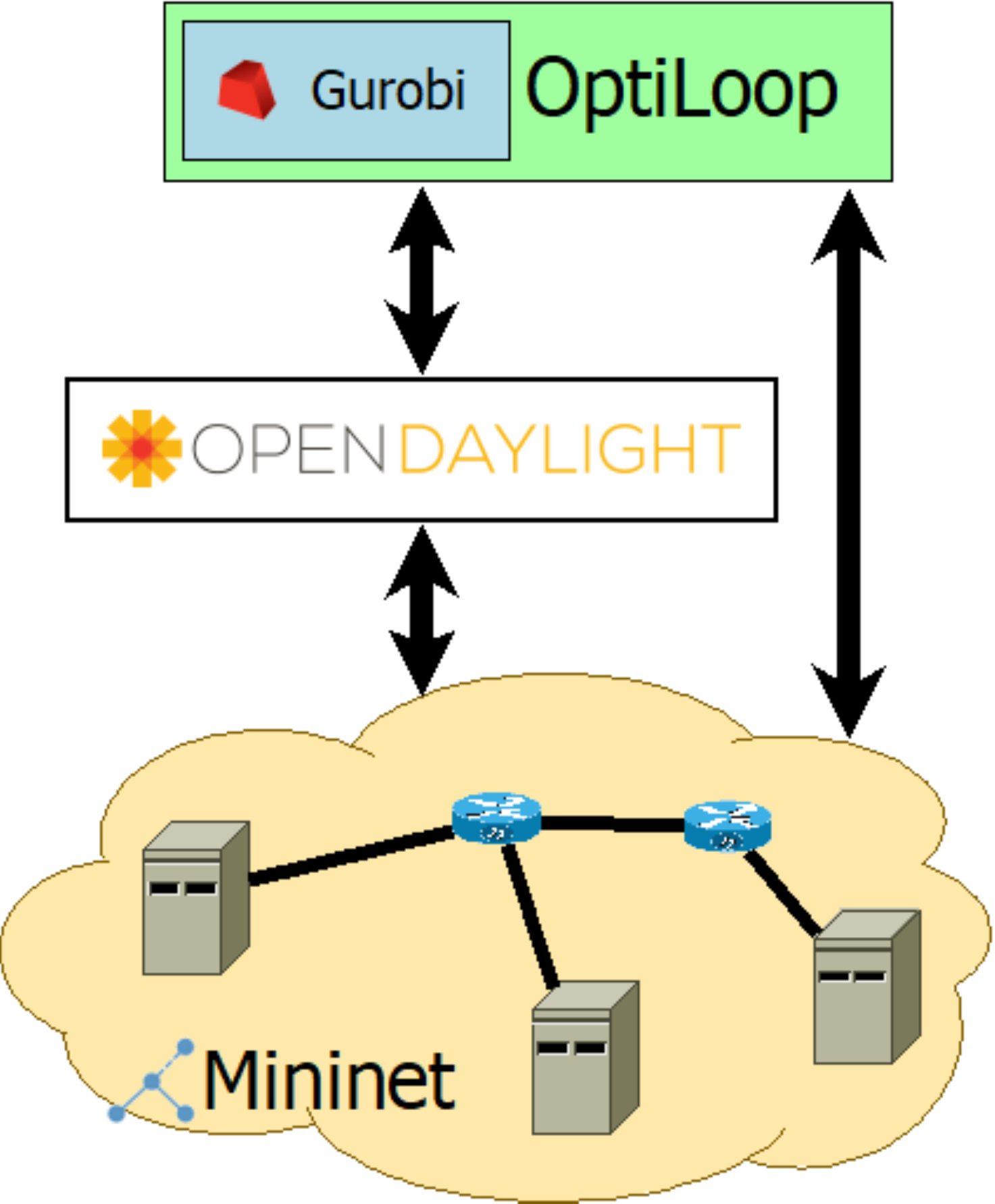}
} 
\hspace{1cm}
\subfigure[\label{fig:mininet-topo}]{
    \includegraphics[trim={0 1.5cm 0 1.5cm},clip,width=1\columnwidth]{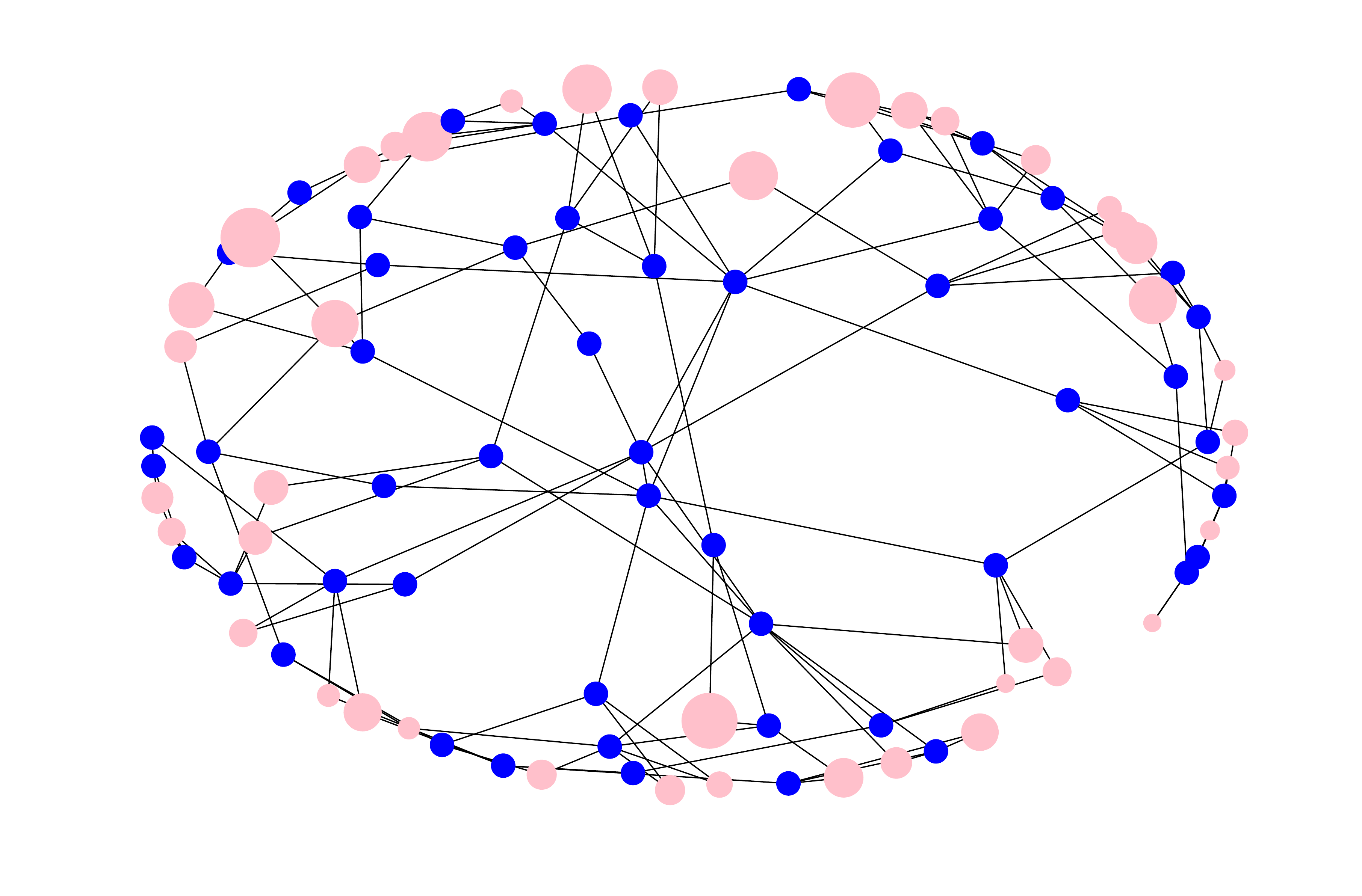}
} 
\caption{
    Architecture (left) and topology (right) of the emulation-based topology. Mininet is used to emulate a network whose topology and traffic match those of a real-world network operator, as discussed in \Sec{topo}.
    The size of pink dots is proportional to the traffic generated by the corresponding endpoint.
    \label{fig:mininet}
} 
\end{figure*}

\begin{figure}
\centering
\includegraphics[width=1\columnwidth]{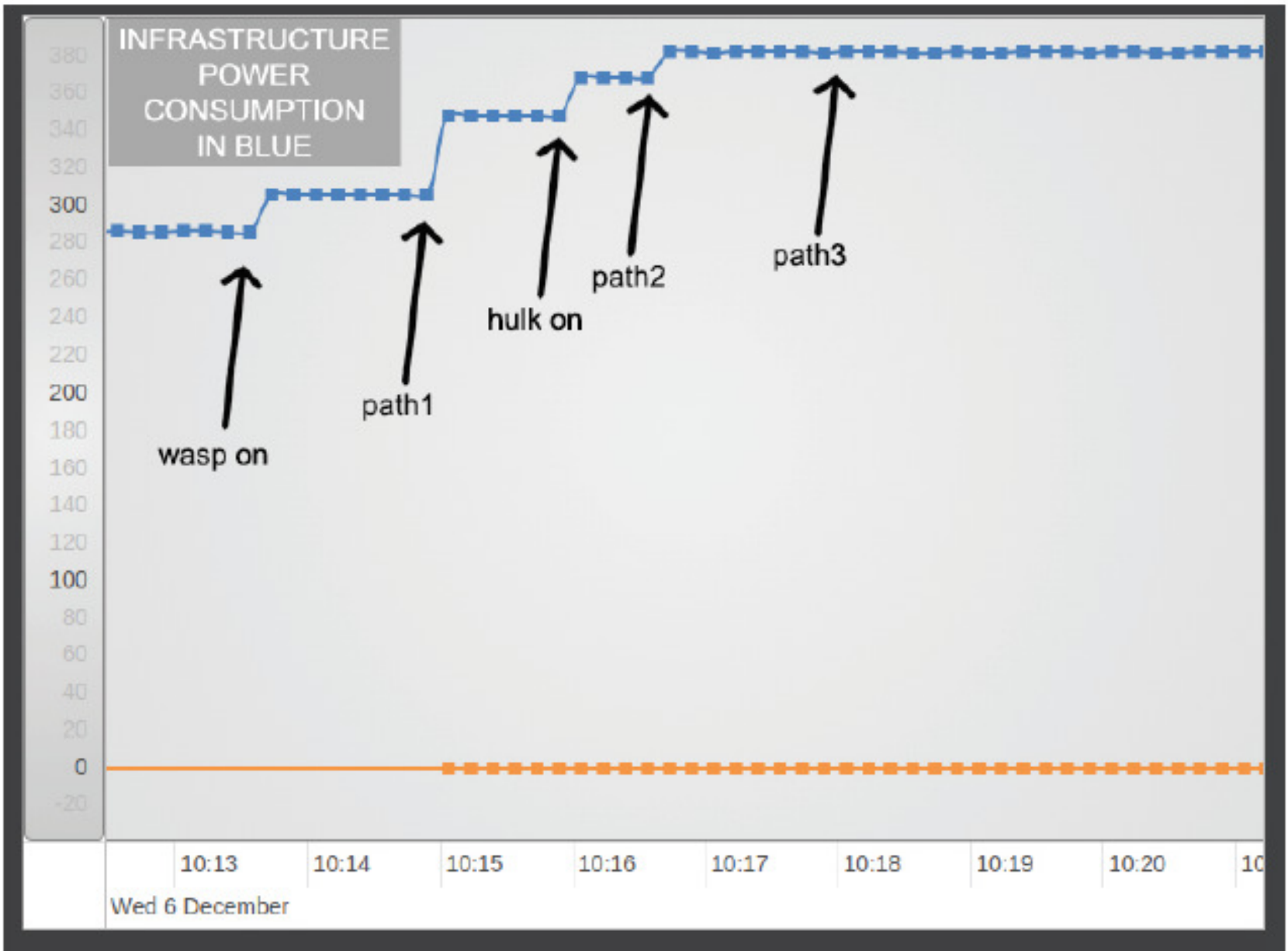}
\caption{Real-world testbed, 
    path instantiation experiment: evolution of the power consumption of the whole network as the three paths are instantiated. Screenshot from the network orchestrator GUI.
    \label{fig:power-evo}
} 
\end{figure}

\begin{table*}[t]
\caption{Real-world testbed,  
path instantiation experiment: power consumption [W]
\label{tab:path-power}
}
\begin{tabularx}{1\textwidth}{|p{4cm}||X|X|X|X|X|X|X|}
\hline
Condition & Switch~$\sharp$1 & Switch~$\sharp$2 & Switch~$\sharp$3 & Switch~$\sharp$4 & Switch~$\sharp$5 & Switch~$\sharp$6 & All switches\\
\hline\hline
All paths off & 21.0299 & 21.0281 & 21.02183 & 20.9614 & 20.9678 & 21.0173 & 125.9841\\
\hline
Path 1 on, no traffic & 35.0349 & 20.9888 & 35.0096 & 21.0168 & 20.9670 & 34.9968 & 167.9023\\
\hline
Path 1 on, with traffic & 35.4876 & 21.0455 & 35.6104 & 20.9996 & 21.0180 & 35.4558 & 168.2835\\
\hline
Paths 1--2 on, no traffic & 35.0309 & 34.9947 & 34.9646 & 20.9988 & 20.9869 & 34.9846 & 181.9242\\
\hline
Paths 1--2 on, with traffic & 35.2771 & 35.2135 & 35.6386 & 21.0171 & 20.9685 & 35.2783 & 182.1381\\
\hline
Paths 1--3 on, no traffic & 34.9826 & 34.9894 & 34.9645 & 20.9861 & 20.9693 & 35.0037 & 181.9220\\
\hline
Paths 1--3 on, with traffic & 35.6249 & 35.7221 & 35.5753 & 21.0042 & 20.9898 & 35.5849 & 183.6007\\
\hline
\end{tabularx}
\end{table*}

\begin{table}[t]
\caption{Real-world testbed, 
provisioning experiment: delays [s]
\label{tab:realtestbed}
} 
\begin{tabularx}{1\columnwidth}{|p{3cm}||X|X|X|}
\hline
Time Component & Maximum & Minimum & Average\\
\hline\hline
OptiLoop & 15.117 & 6.144 & 9.729\\
\hline
Server activation & 0.111 & 0.068 & 0.086\\
\hline
Switch activation & 2.660 & 0.871 & 1.824\\
\hline
Virtual links creation & 31.797 & 21.953 & 27.369\\
\hline
Single VNF creation & 38.823 & 30.235 & 31.476\\
\hline
Creation of all VNFs & 49.738 & 31.883 & 38.384\\
\hline
Network path setup & 0.065 & 0.050 & 0.056\\
\hline
Single VNF configuration & 187.199 & 53.854 & 105.860\\
\hline
Configuration of all VNFs & 316.992 & 86.885 & 216.591\\
\hline
Total NS instantiation & 404.868 & 164.036 & 299.522\\
\hline
\end{tabularx}
\end{table}


\subsection{Computational complexity}
\label{sec:complexity}

The \path{fixProblems} and \path{saveEnergy} procedures are run in order to react to changes in the network load; therefore, it is important that the decisions they make are {\em swift} as well as effective. To this end, we can prove that both procedures have polynomial worst-case computational complexity, as stated by the following theorem:
\begin{theorem}
\label{thm:poly}
The \path{fixProblems} (\Alg{fixproblems}) and \path{saveEnergy} (\Alg{saveenergy}) procedures have polynomial worst-case computational complexity.
\end{theorem}
\begin{IEEEproof}
The proof follows by inspection of \Alg{fixproblems} and \Alg{saveenergy}. The algorithms contain no loops, i.e., each of the instructions therein is executed at most once. Among the instruction, all perform elementary operations, except:
\begin{itemize}
    \item finding the minimum of a set, which requires sorting and has complexity~$O(n\log n)$, $n$ being the set size;
    \item solving convex optimization problems, which has polynomial, namely, cubic computational complexity~\cite{boyd}.
\end{itemize}
Thus, the overall complexity of the \path{fixProblems} and \path{saveEnergy} procedures is polynomial, namely, cubic.
\end{IEEEproof}

\Thm{poly} ensures that the \path{fixProblems} and \path{saveEnergy} can be used to make swift and effective decisions in reaction to traffic changes. Indeed, convex optimization problems are routinely~\cite{boyd} solved in embedded applications with real-time requirements.

\section{Testbeds, scenario and benchmarks}
\label{sec:scenario}

We validate and evaluate OptiLoop through two testbeds. We study the interaction between OptiLoop, the SDN controller, and the  NFVO in a small-scale testbed with real hardware, described in \Sec{realtestbed}. For our performance evaluation we instead use a larger, emulated testbed based on the real-world topology of a mobile operator, as detailed in \Sec{mininet}. 
In all experiments, the reference VNF graph is the vEPC service described in \Fig{logical}.

\subsection{Real-world testbed}
\label{sec:realtestbed}

The architecture and topology of our real-world testbed are described in \Fig{realtestbed}. OpenDaylight (Beryllium version) and OpenStack (Mitaka version) are used to control a network made of six Lagopus software switches (with DPDK support enabled for faster switching) and three physical servers, connected as shown in \Fig{real-topo}. The OpenDaylight SDN controller configures the data plane, by activating/deactivating links and switches via SNMP protocol and configuring the forwarding rules via OpenFlow 1.3 protocol. A custom-built NFVO -- integrated with the VNFM (VNF manager) and VIM (Virtual Interface Manager) OpenStack modules -- manages the VMs that run the VNFs. 
Specifically, the NFVO provides RESTful interfaces that allows the orchestration of network services. Services themselves are is composed by multiple VNFs, which are interconnected through the specification of a VNF graph. 
A detailed description of its architecture and implementation can be found in~\cite[Sec.~2.6]{xhaul32}. 
We adopt the OpenAirInterface~\cite{openair} vEPC implementation, including the four VNFs in \Fig{logical}.

OptiLoop is implemented as a standalone application, written in Java and including two main components, devoted to monitoring and decision-making. OptiLoop interacts with both OpenDaylight and the NFVO through their REST APIs, gathering up-to-date information on the status of switches, links, physical servers and VNFs. When a decision is made, it communicates it to OpenDaylight (if the decision concerns link activation/deactivation) or the NFVO (if the decision concerns VNF deployment or server activation/deactivation). The decision-making component essentially implements \Alg{fixproblems} and \Alg{saveenergy}, using the Gurobi solver for optimization.  Since Gurobi features Java bindings, using it within the OptiLoop application is as simple as importing a library.

\subsection{Emulated testbed}
\label{sec:mininet}

Our performance evaluation is carried out through an emulated testbed based on Mininet, the {\em de facto} standard solution to study SDN-based networks. Its architecture is summarized in \Fig{mininet-archi}: similarly to the previous case, OptiLoop interacts with the OpenDaylight controller for network management, and directly with Mininet via its Python API to turn servers and switches on and off. Notice that the actual VNFs are {\em not} implemented in Mininet; the traffic they serve is emulated via \path{iperf} and the energy consumption is estimated from our real-world testbed, as detailed in \Sec{exp-path} next.

The switches and servers emulated by Mininet reproduce the real-world topology of a major mobile operator, as detailed in \Sec{topo}. Links and servers are implemented through the \path{TCLink} and \path{CPULimitedHost} Mininet classes, which allow us to assign them bandwidth, delay and computational capability matching those of their real-world counterparts. All \path{iperf}-generated traffic is based on the real-world traffic figures we have access to.

\subsubsection{Network topology and traffic}
\label{sec:topo}

Our reference topology, displayed in \Fig{mininet-topo},  represents the real-world topology of a major mobile network operator. It includes 42~endpoints and 51~B/F nodes, with each endpoint connected to exactly two B/F nodes. A total of 1,497 antennas are connected to the endpoints. 
In the trace, per-endpoint traffic varies between \SI{23.3}{\mega bit/\second} and \SI{148.9}{\mega bit/\second}. In order to model future network conditions, we increase such values by accounting for the 22\% annual growth rate foreseen by Cisco~\cite{cisco} for the next five years, thus obtaining per-endpoint traffic values  
varying between \SI{74}{\mega bit/\second} and \SI{473}{\mega bit/\second} per endpoint, with a 82:18 downlink/uplink proportion. The dataset we use only represents a snapshot of the network conditions, i.e., traffic demand does not change over time.

Based on the real-world vEPC implementation~\cite{openair} we consider a total of four VNFs, namely eNB, MME, HSS, and a gateway  implementing both the P-GW and S-GW functions. Notice that in \cite{openair} no VNF is split into user- and control-plane sub-entities. We set our $\chi$-values, expressing how traffic gets transformed as it travels between VNFs, leveraging the analysis in~\cite{globecom15_bearer}; in particular, the fraction of control traffic going to the MME is given by~$\chi(\text{eNB},\text{P/S-GW},\text{MME})=0.32$.

Still based on~\cite{globecom15_bearer}, we set the link bandwidth~$B_{i,j}$ to \SI{10}{\giga bit/\second} for endpoint-to-node links and \SI{100}{\giga bit/\second} for node-to-node ones. Based on~\cite{globecom15_bearer} and~\cite{lagopus-fun}, we assume that each B/F node can process \SI{100}{\giga bit} of traffic every second. Since our scenario is constrained by B/F node and link capacity, we ignore network and processing delays.

\begin{figure*}
\centering
\includegraphics[width=.32\textwidth]{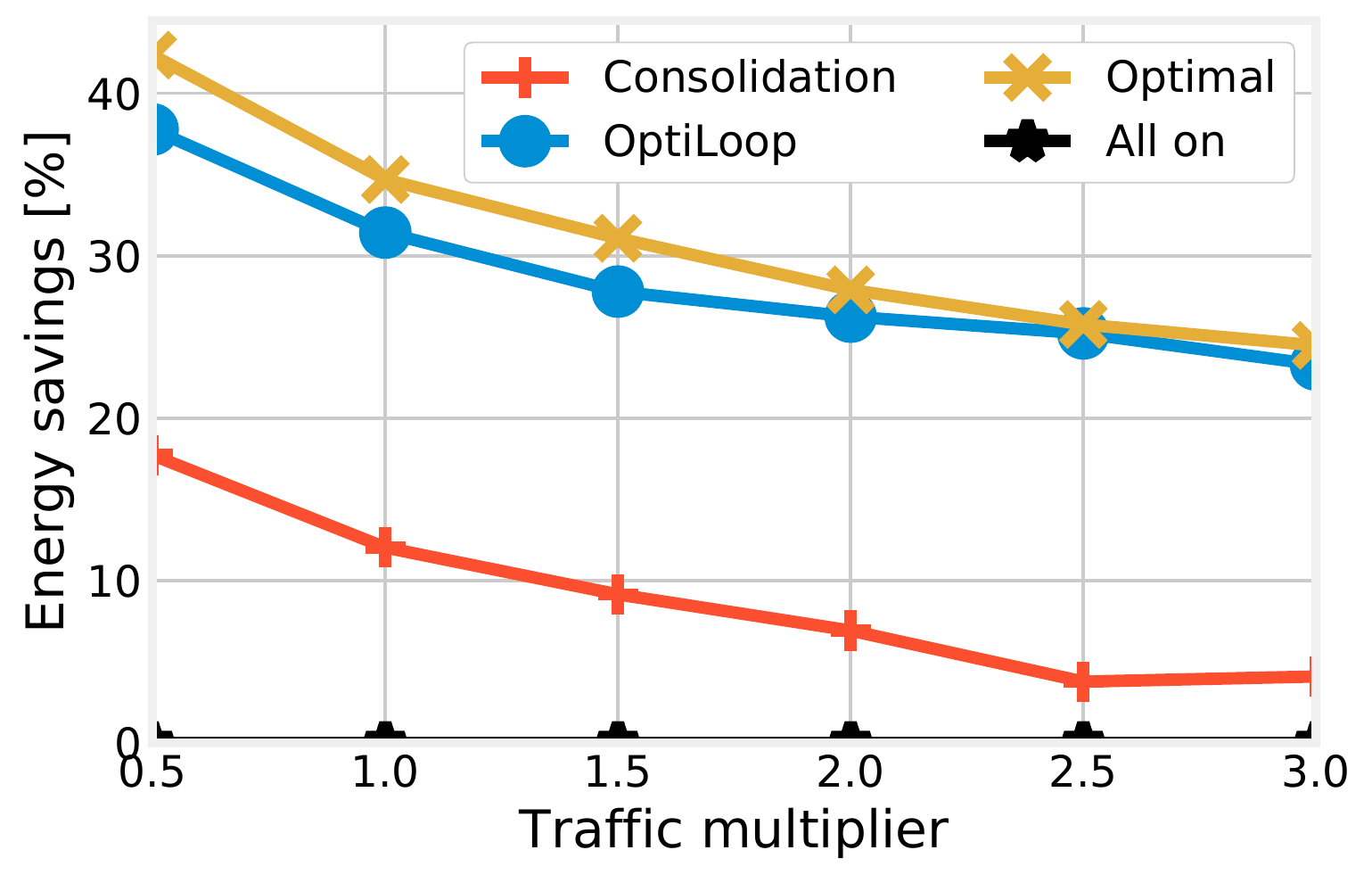}
\includegraphics[width=.32\textwidth]{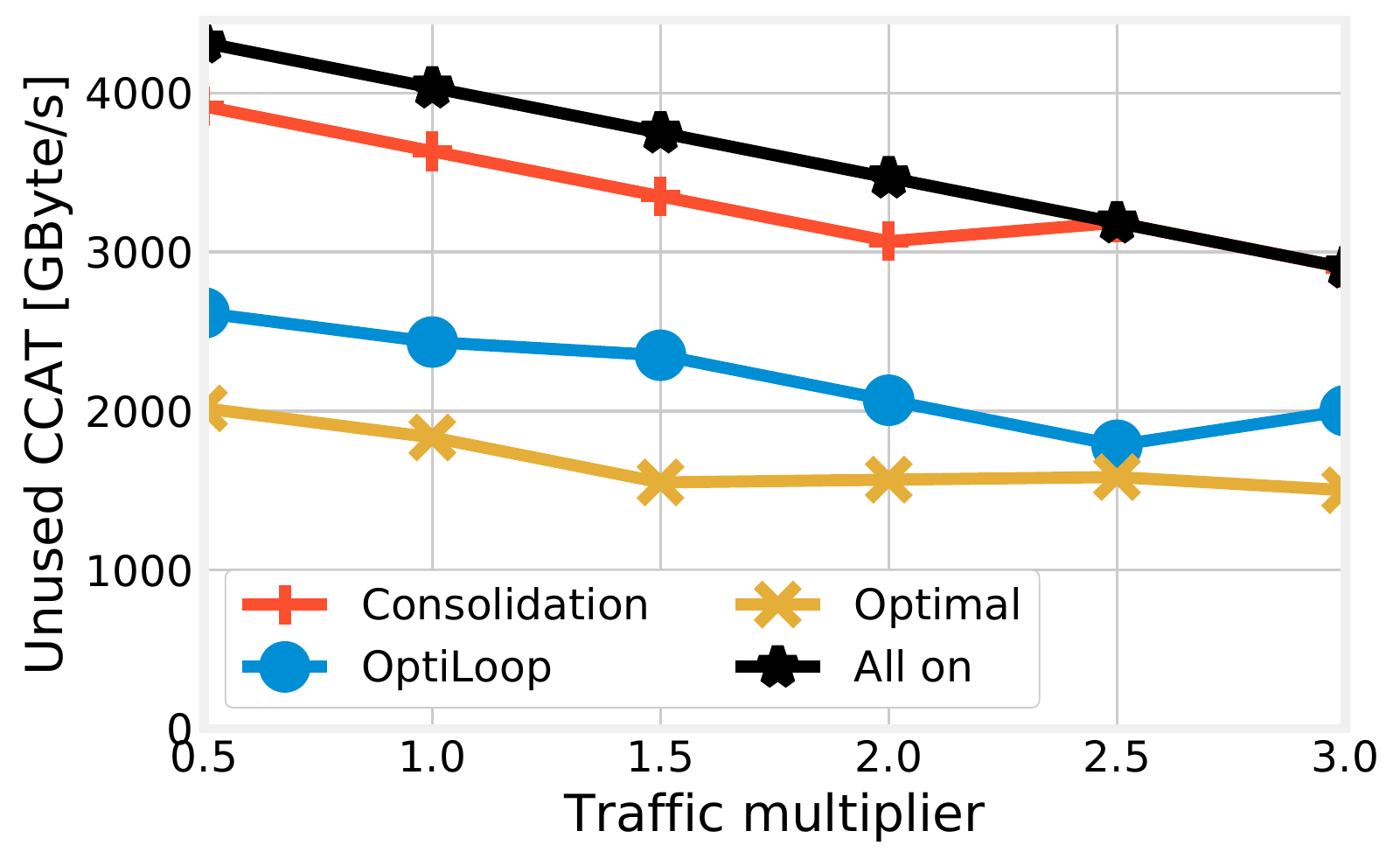}
\includegraphics[width=.32\textwidth]{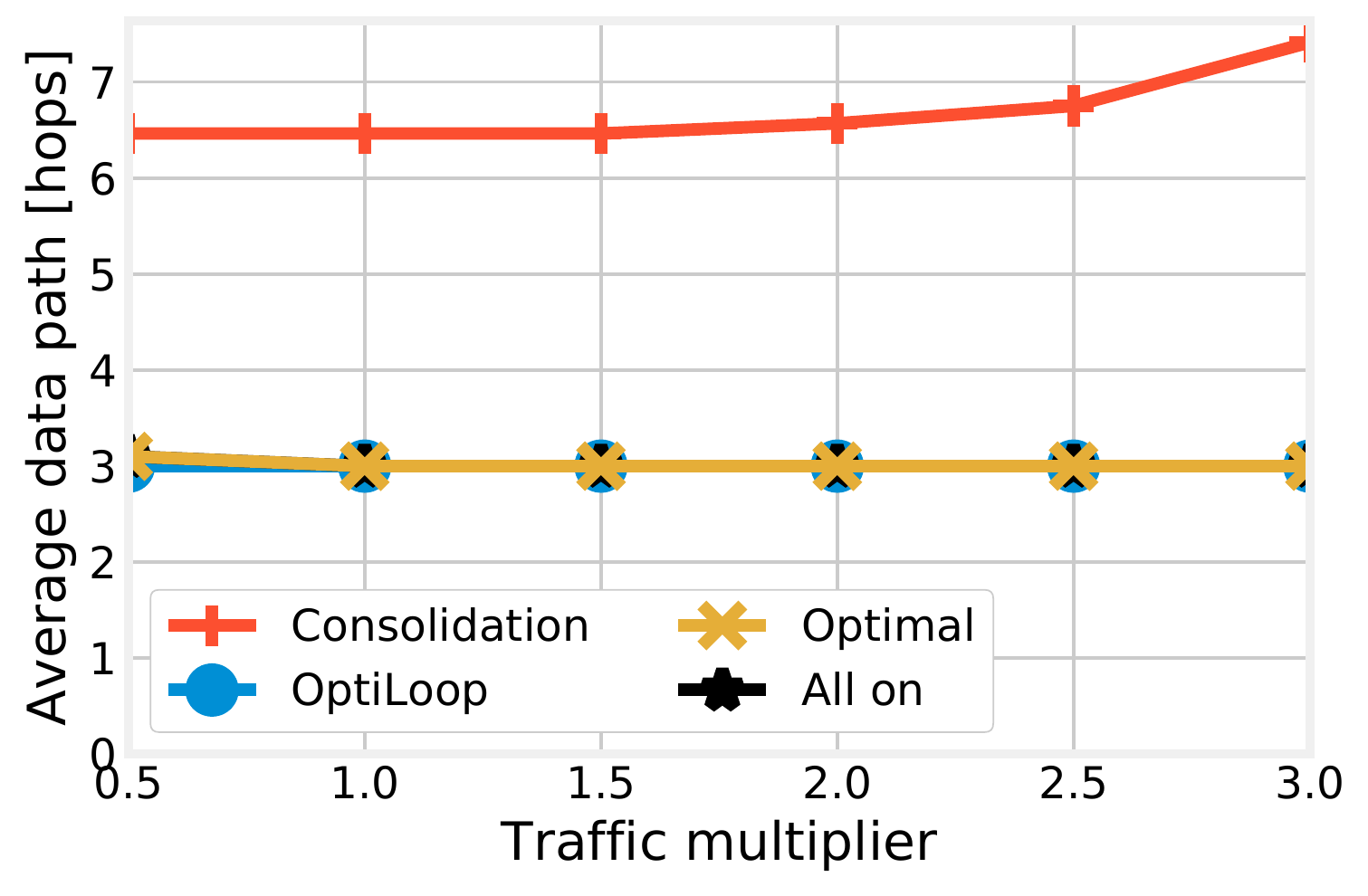}
\caption{Mininet experiments with real-world topology: 
energy savings obtained as a function of traffic (left); spare computational capabilities of the active topology (CCAT) (center); number of hops traveled by requests (right).
    \label{fig:general}
}
\end{figure*}
\begin{figure}
\centering
\includegraphics[width=.8\columnwidth]{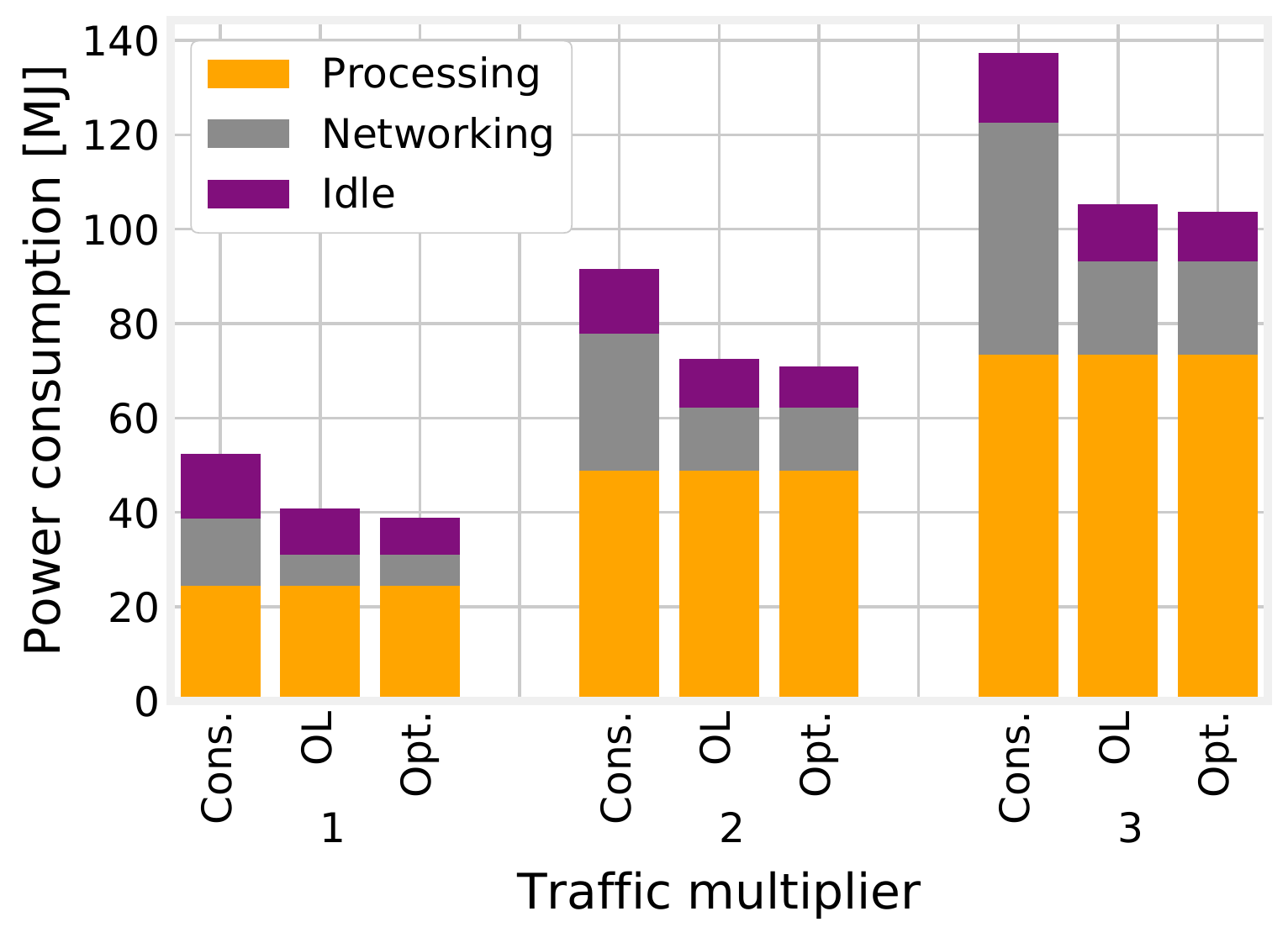}
\caption{Mininet experiments with real-world topology: 
breakdown of energy consumption for the consolidation-based (``Cons.''), OptiLoop (``OL''), and optimal (``Opt.'') strategies.
    \label{fig:breakdown}
}
\end{figure}

\subsubsection{Benchmark solutions}
\label{sec:benchmarks}

We compare OptiLoop with 
three alternatives:
\begin{itemize}
    \item what is done in real-world systems, i.e., keeping all network elements active regardless of traffic, indicated as {\em All on} in the plots;
    \item the optimal solution obtained by brute-force, i.e., trying all possible combinations of network elements to activate, indicated as {\em Optimal} in the plots;
    \item a state-of-the-art approach based on consolidation, based on~\cite{cloudnet16_energyaware} and indicated as {\em Consolidation} in the plots.
\end{itemize}

The consolidation procedure used in~\cite{cloudnet16_energyaware} consists of three-stage decision process. For every flow, it first looks for an already-deployed VNF to serve the flow; if none can be found, it deploys a new instance of the  VNF at an already active B/F node. If no suitable node is found, it activates a new one. Also, the procedure activates any additional B/F nodes needed to ensure connectivity between endpoints and the serving B/F nodes.
It is interesting to notice how all stages of the consolidation design process have the same goals of our \path{fixProblems} procedure, namely, ensuring that there is enough computational capability (steps~1 and~2) and network capacity (step~3) to process the incoming traffic. There is no equivalent for the \path{saveEnergy} procedure, i.e., already-made decisions are never reconsidered.


\begin{figure*}
\centering
\includegraphics[width=.26\textwidth]{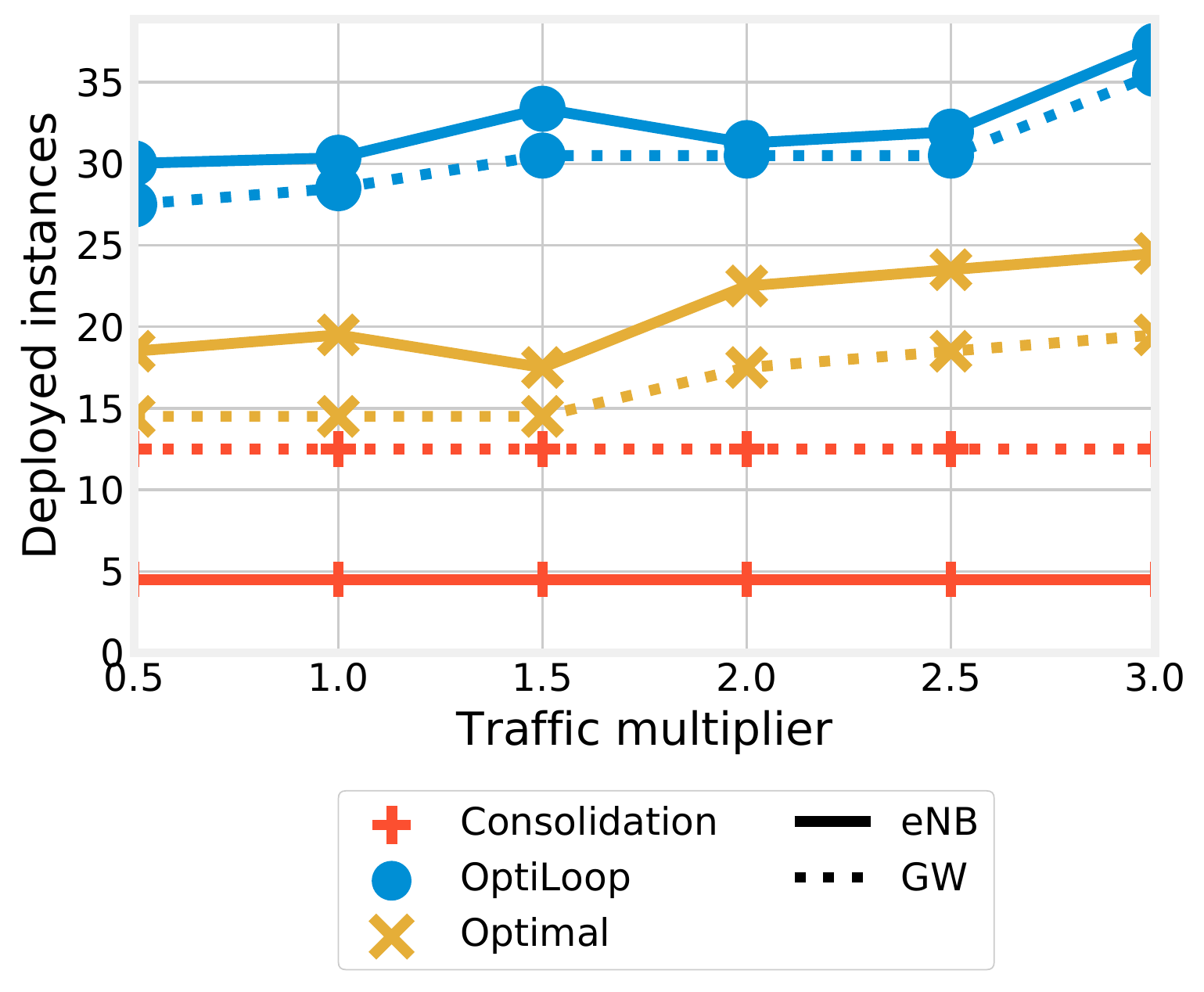}
\includegraphics[width=.32\textwidth]{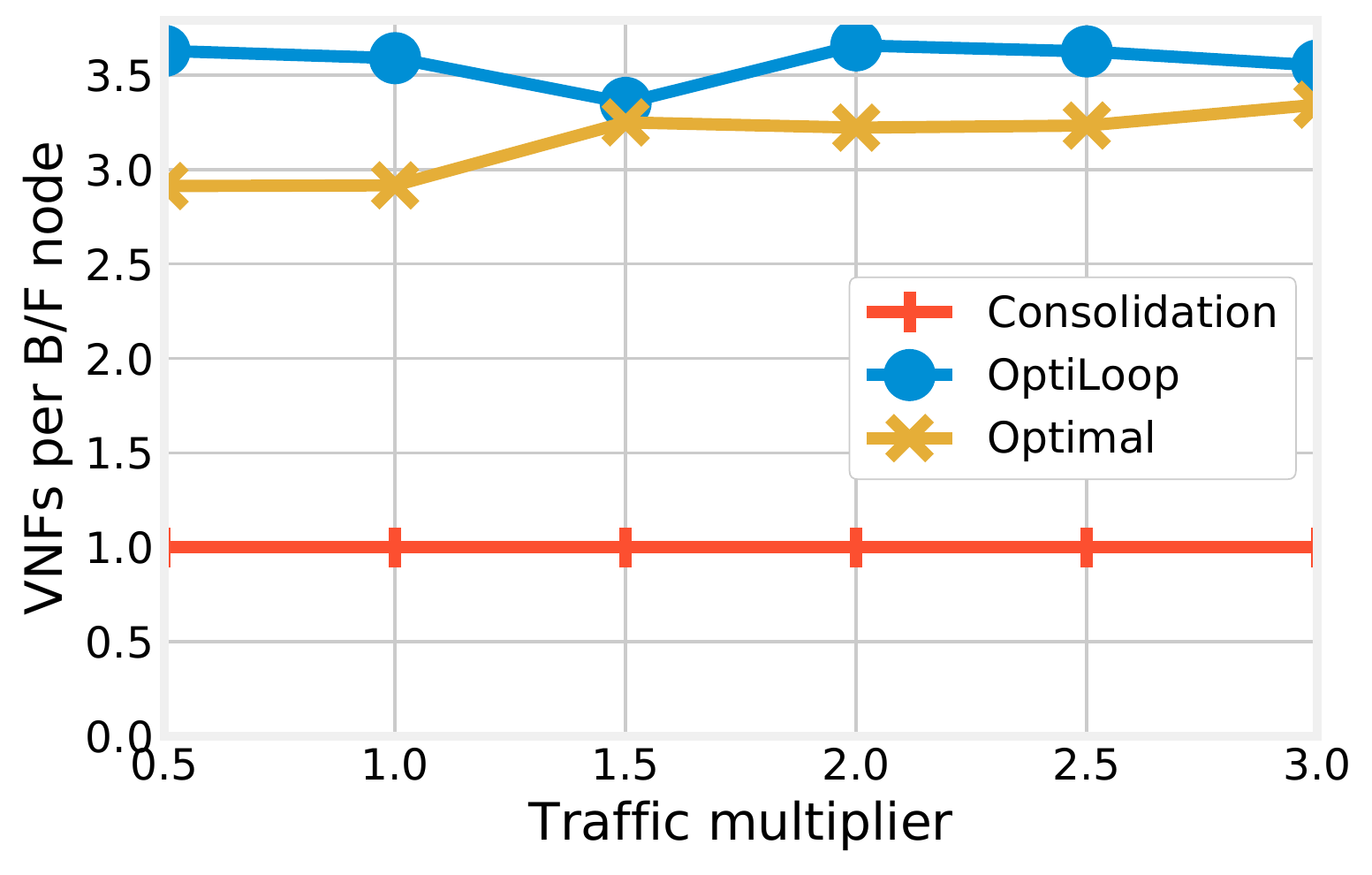}
\includegraphics[width=.32\textwidth]{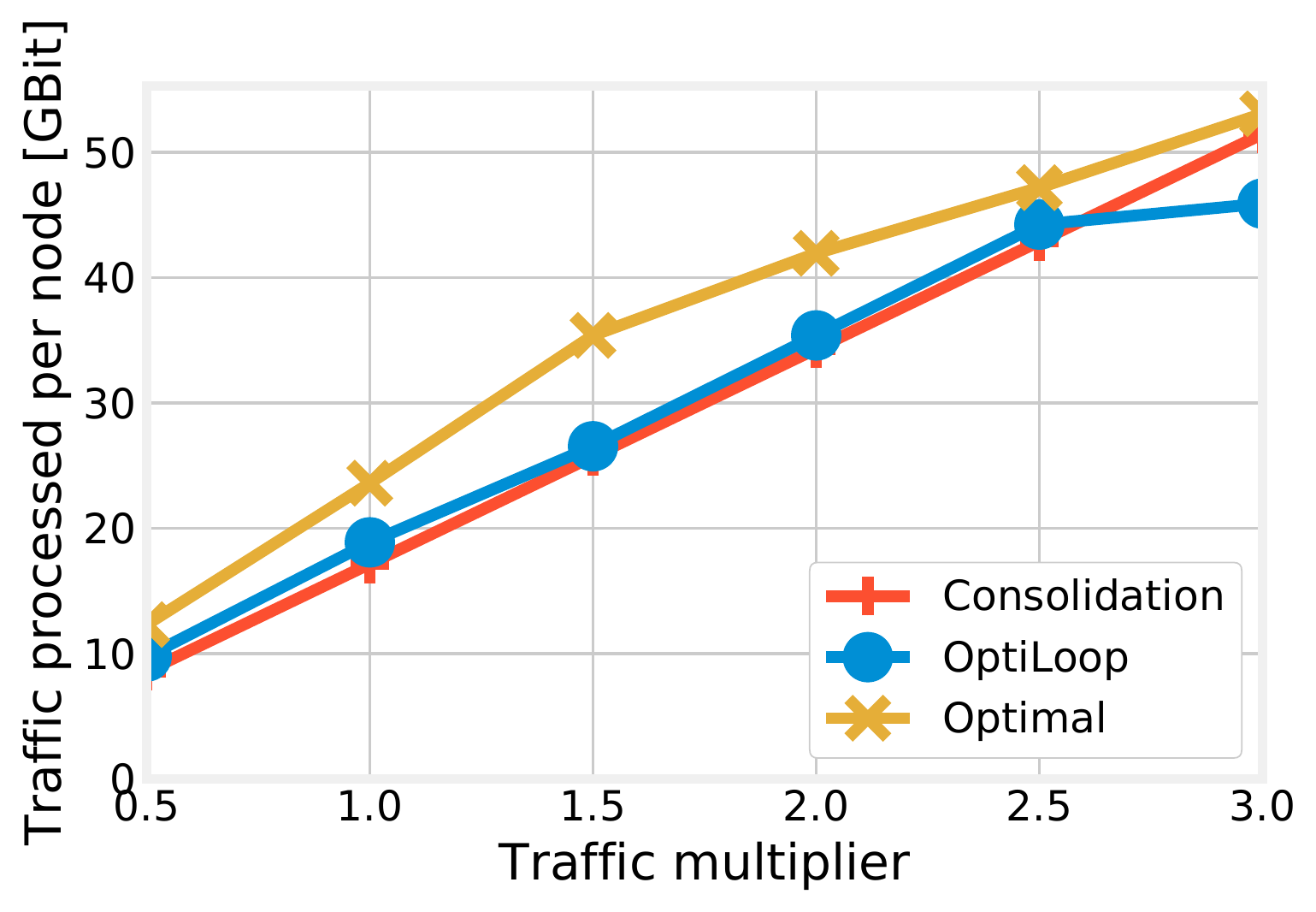}
\caption{Mininet experiments with real-world topology: 
number of deployed instances for the eNB and P/S-GW VNFs (left); average number of VNFs deployed in each B/F node (center); average traffic processed at each B/F node (right).
    \label{fig:details}
}
\end{figure*}

\section{Results}
\label{sec:results}

We start this section by summarizing,
in \Sec{res-validation}, the power consumption and delay figures we obtain from the real-world testbed described in \Sec{realtestbed}.
We then present, in \Sec{res-emulation}, a performance evaluation of OptiLoop carried out by emulating a real-world topology in Mininet, as described in \Sec{mininet}.

\begin{figure}
\centering
    \includegraphics[trim={0cm 1.5cm 0cm 1.5cm},clip,width=1\columnwidth]{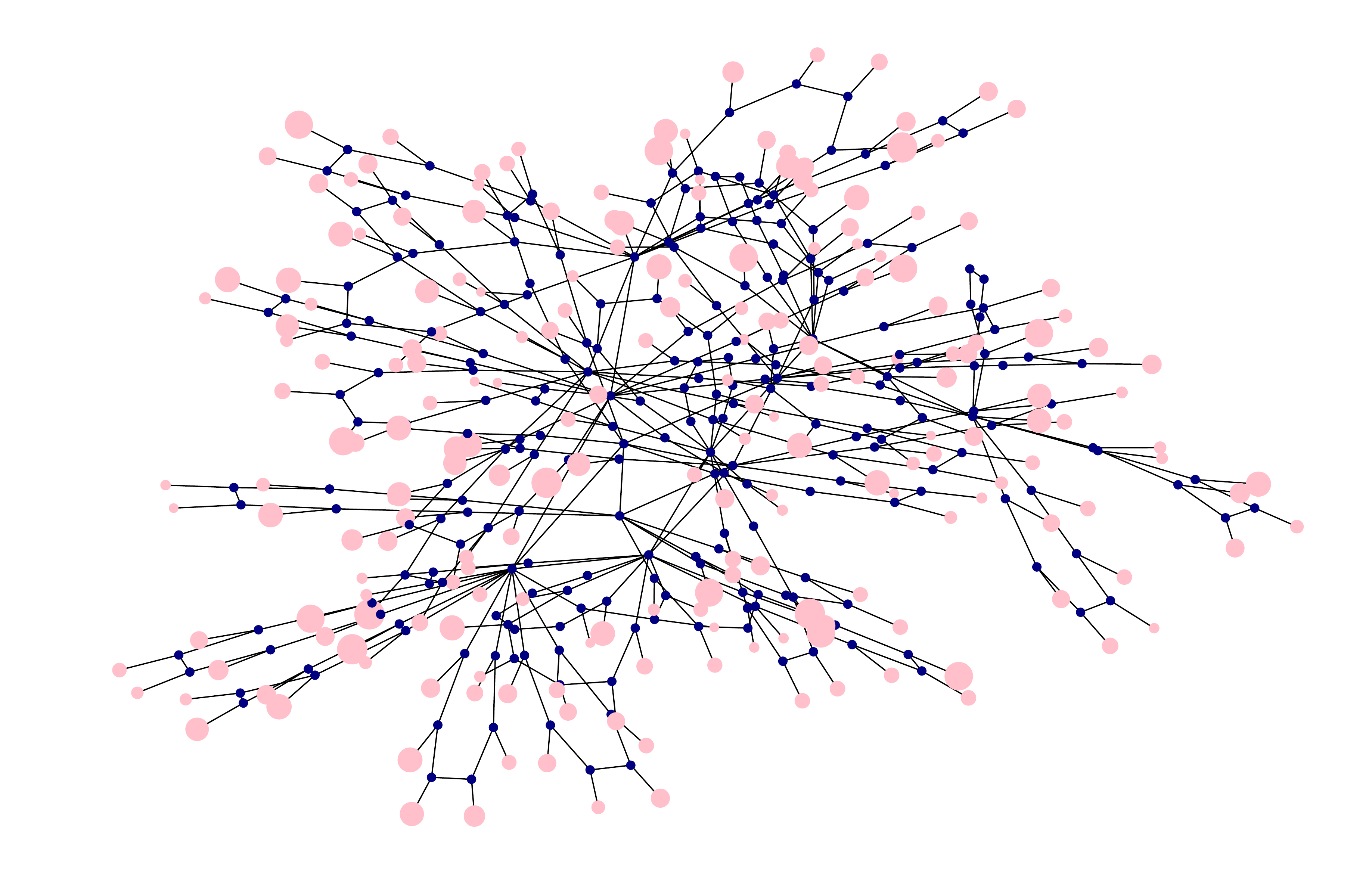}
\caption{
Scaled-up network topology. As in \Fig{mininet}, blue dots indicate B/F nodes, pink ones indicate endpoints, and the size of pink dots is proportional to the traffic generated by the corresponding endpoint.
\label{fig:topo-big}
} 
\end{figure}

\subsection{Results from the real-world testbed}
\label{sec:res-validation}

There are two main types of information we seek to obtain from the real-world testbed described in \Sec{realtestbed}:
\begin{itemize}
    \item the {\em power consumption} associated with B/F nodes, broken down in idle and processing power;
    \item the {\em delay} associated with changes to the network, e.g., activating a link or instantiating a new VM.
\end{itemize}
We measure the above quantities through two experiments, namely, a path instantiation experiment and a service provisioning one, as described next.

\subsubsection{Path instantiation experiment}
\label{sec:exp-path}

In this experiment, we start with all equipment -- switches and servers -- in sleeping mode. We then instantiate, one by one, the three paths shown in \Fig{real-topo}, activating additional switches as needed. Finally, we generate bidirectional flows of \SI{1}{\giga byte/\s} between each pair of endpoints, so as to ascertain the impact of traffic on the power consumption.

The evolution of the power consumption in our real-world testbed is exemplified in \Fig{power-evo}. In the beginning, when all network elements are in sleeping mode, the total power consumption is around \SI{280}{W}. Activating new servers results in an increase in power consumption, as can be expected. More interestingly, instantiating a new path results in a power increase only if it requires activating a new switch, as is the case of path~1 and path~2. As we can see from \Fig{real-topo}, path~3 requires no extra switches with respect to path~1 and path~2, and therefore instantiating it results in no additional consumption.

\Tab{path-power} provides a more analytical view of the power consumed by the switches in different states. When all equipment is in sleeping mode (first row), each switch consumes roughly \SI{21}{W} of power. Instantiating path~1 (second row) requires activating switches~1--3 and~6, whose power consumption jumps to \SI{35}{W}; activating additional paths has the same effect on the other switches. We can also observe that sending traffic over the instantiated paths has a noticeable, but minor, effect: routing \SI{1}{\giga byte/s} of traffic results in an additional consumption of around \SI{0.5}{W} per switch. Finally, notice that the last column of \Tab{path-power} does not match the line in \Fig{power-evo} since the latter also includes the consumption of the physical servers, i.e., \SI{80}{W} in sleeping mode and roughly \SI{120}{W} when active.

\subsubsection{Service provisioning experiment}
\label{sec:exp-vm}

In the service provision experiment, we are interested in measuring the {\em delay} associated with performing changes to the network, including path instantiation and service provisioning. To this end, we use the network described in \Sec{realtestbed} to provide the virtual EPC (vEPC) service, consisting of the VNFs depicted in \Fig{logical}, as implemented in~\cite{openair}.

Doing so requires three main steps, namely (i) making VNF placement and traffic routing decisions, i.e., running OptiLoop; (ii) setting up the required paths, similar to the path instantiation experiment described in \Sec{exp-path}; (iii) instantiating and configuring the VMs that run the VNFs. The aspect we are chiefly interested in is the relative importance of such delay components.
The results are summarized in \Tab{realtestbed}. A first, important observation is that OptiLoop only accounts for a small fraction (roughly~3\%) of the total delay; in other words, the energy savings it brings come at a modest price in terms of additional delay.

Among the other delay components, we can observe that VM configuration and, to a lesser extent, virtual link creation dominate the total delay. It is also interesting to notice the values labeled ``Creation of all VNFs'' and ``Configuration of all VNFs'', which are substantially less than four times the creation (resp. configuration) of a single VNF. This is because, once decisions are made by OptiLoop, they can be implemented in a parallel fashion.

\begin{figure*}
\centering
\includegraphics[width=.32\textwidth]{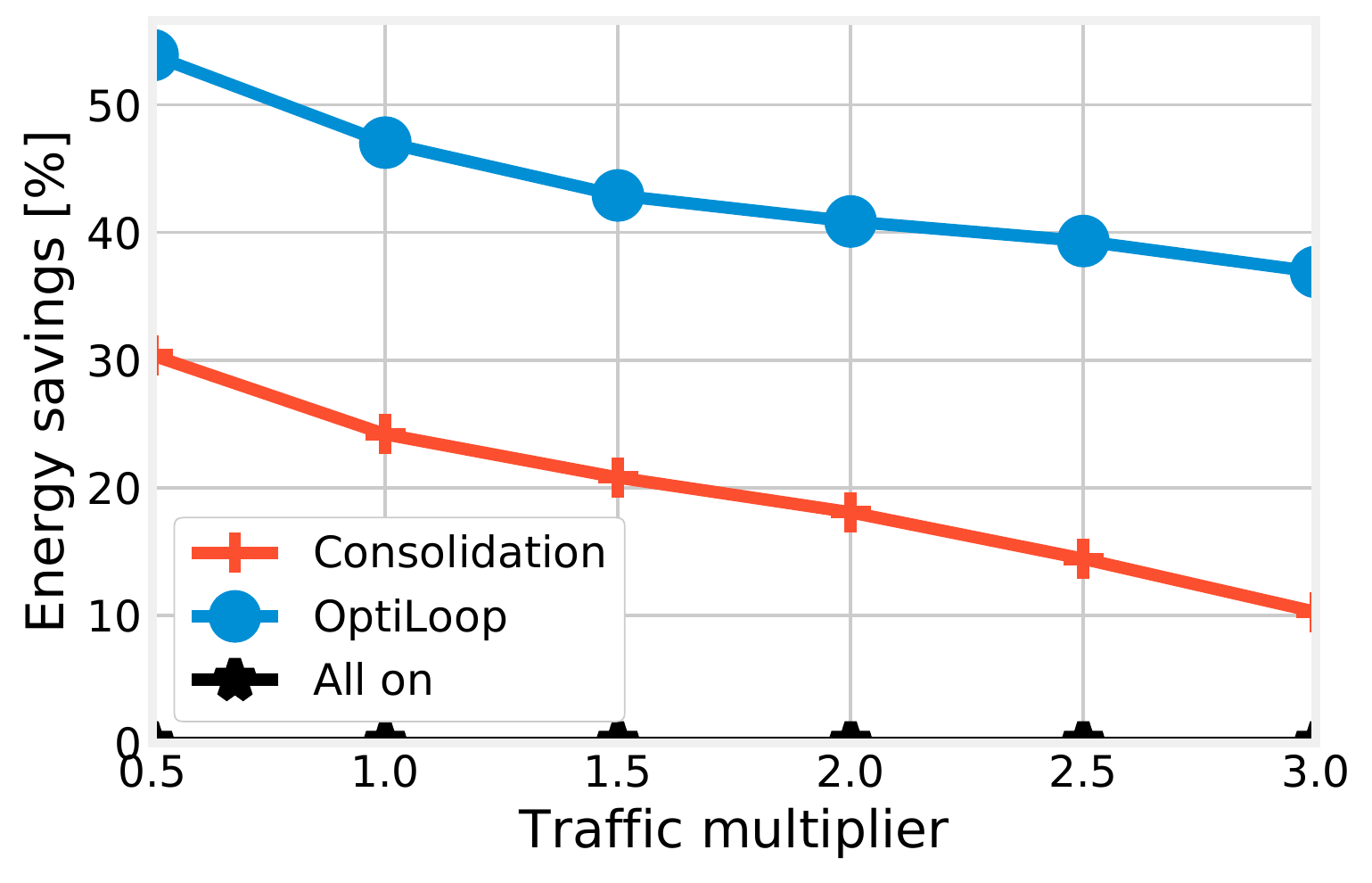}
\includegraphics[width=.32\textwidth]{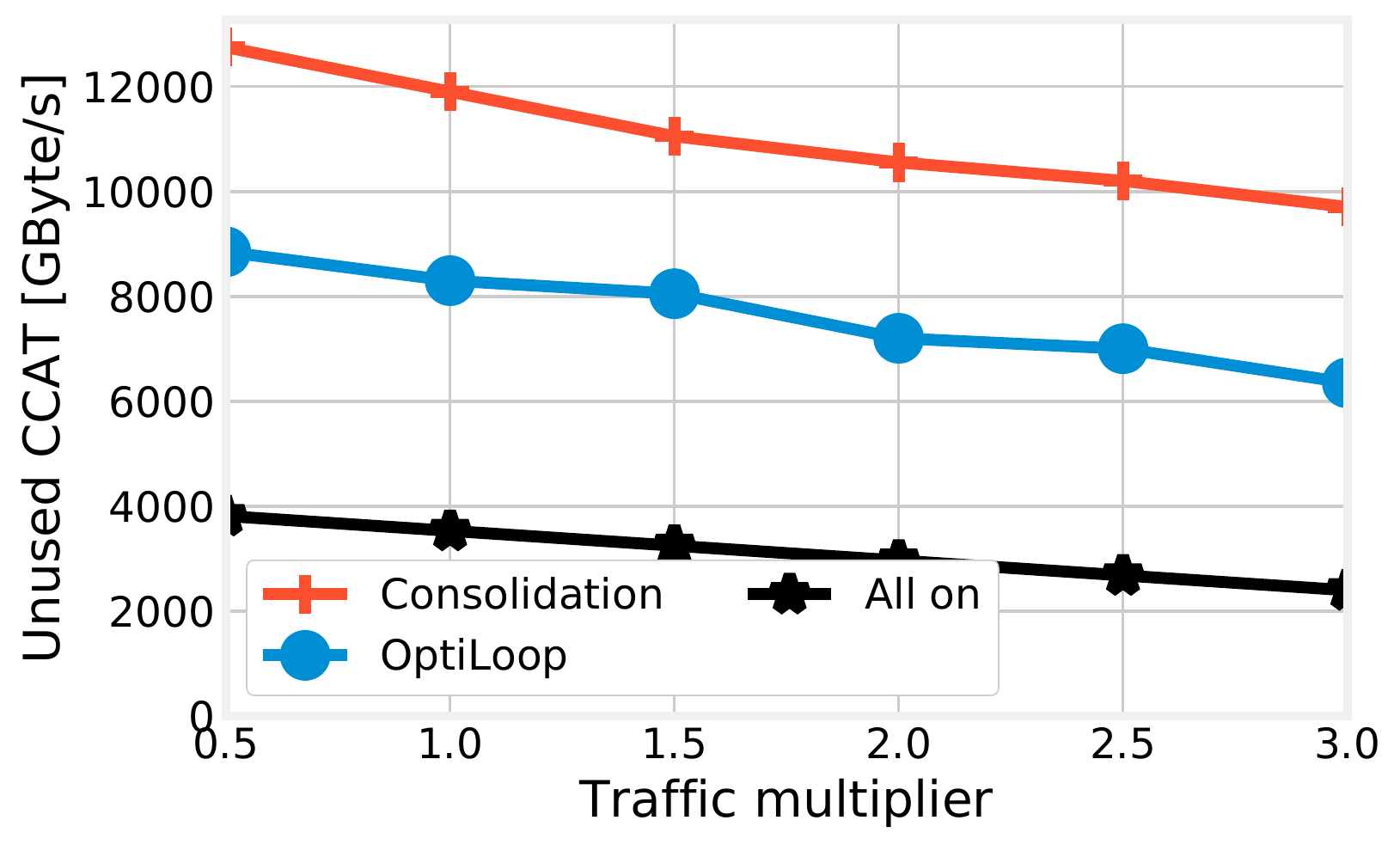}
\includegraphics[width=.32\textwidth]{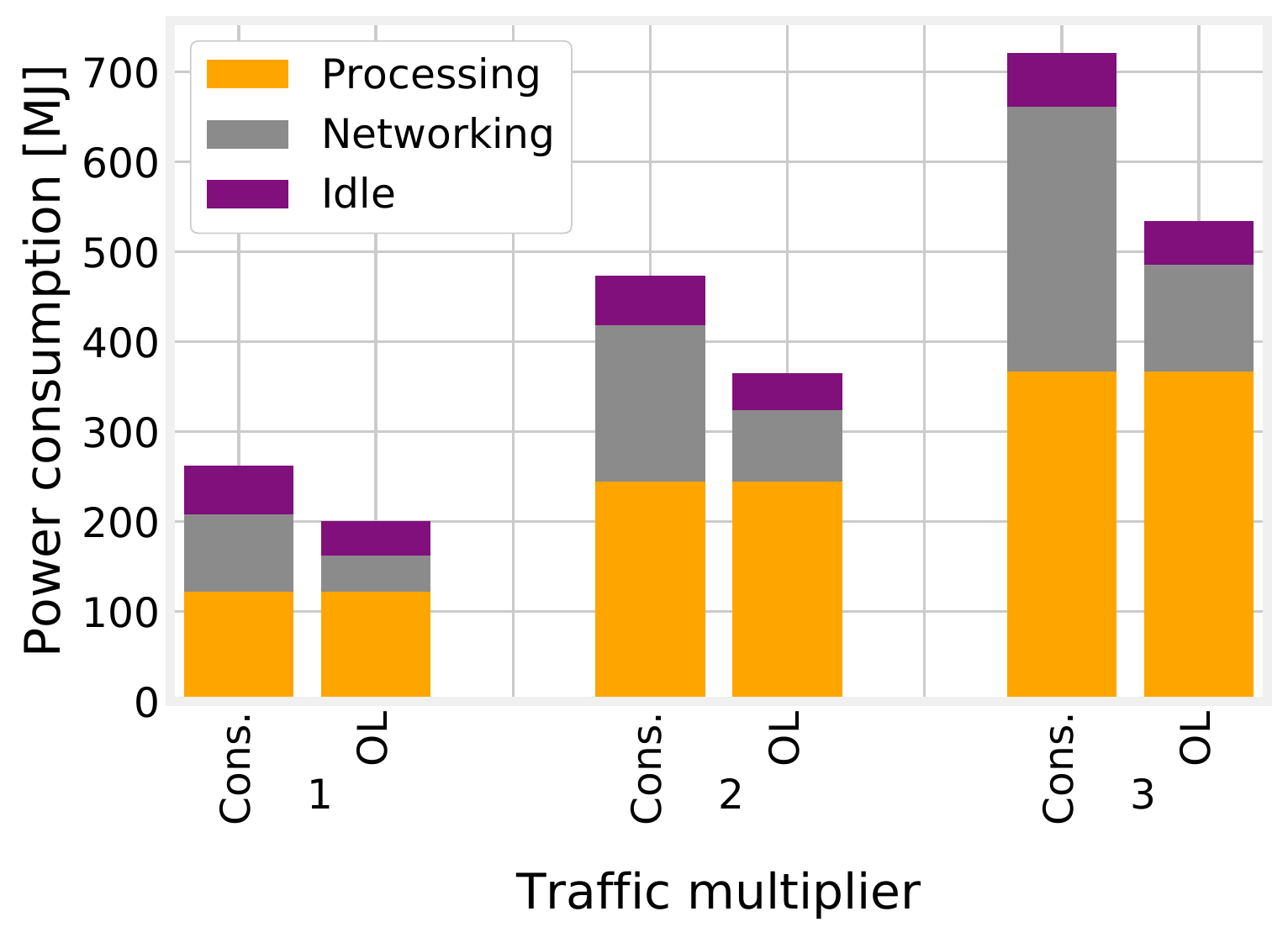}
\caption{Mininet experiments with scaled-up topology: 
savings obtained as a function of traffic (left); spare computational capabilities of the active topology (CCAT) (center); energy consumption breakdown (right). Traffic multipliers are referred to the scaled-up traffic, i.e., five times the traffic in the original trace described in \Sec{topo}.
    \label{fig:scaledup}
}
\end{figure*}

\subsection{Emulation-based performance evaluation}
\label{sec:res-emulation}

The first answer we seek from the performance evaluation carried out through the emulated testbed concerns the magnitude of possible energy savings. In \Fig{general}(left), we vary the traffic demand between~0.5 and 3~times the real-world amount, and study how much energy we can save compared to what is done today, i.e., leaving all B/F nodes and links active. We can observe that OptiLoop yields dramatic savings, consistently very close to the optimum, while consolidation does not perform as well. An intuitive reason  is that OptiLoop accounts for all the three main contributions to energy consumption (processing, idle power, and networking), while the consolidation-based approach focuses on keeping the number of active B/F nodes low.

\Fig{general}(center) shows the spare computational capability of the active topology (CCAT); intuitively, this is a measure of how much power is being wasted, i.e., how inefficient the network management strategy is. The consolidation algorithm has the highest spare CCAT, because of the higher number of B/F nodes that have to be activated in order to guarantee connectivity. The spare CCAT yielded by OptiLoop is much lower, and very close to the optimum. It is interesting to remark that even the optimum leaves substantial spare CCAT. This is due to the fact that some B/F nodes have to be active in order to keep the topology connected, even if they do not have to host any VNF. 
\Fig{general}(right) depicts how many hops data travels across the network. OptiLoop again matches the optimum, while the consolidation strategy results in substantially longer paths, due to the fact that VNF placement decisions are made without accounting for connectivity.

We now use the power consumption we measure from our real-world testbed (\Sec{res-validation}) to extrapolate the total power that the emulated network would consume. \Fig{breakdown} breaks such a consumption into its main components, namely, processing, networking, and idle power.
Note that these components have comparable magnitude, i.e., none of them dominates the overall consumption. It follows that network management strategies have to account for them all. We can also see that the processing component never changes across strategies, since the amount of traffic to process is always the same. The difference between the strategies lies mostly in the networking component (longer paths in \Fig{general}(right) correspond to higher consumption) and, to a lesser extent, in the idle energy. In other words, it is important to  place VNFs close to the traffic they have to serve, while at the same time activating as few B/F nodes as possible.

Dropping the ``all on'' strategy to keep plots easy to read, \Fig{details}(left) and \Fig{details}(center) show that placing VNFs close to the traffic they serve also means placing {\em many} of them. This goes against the traditional concept of activating only the strictly required number of elements, and it is a direct consequence of the features of modern, software-based networks. Indeed, there is little or no penalty for placing an underutilized VNF instance on an already active B/F node, while there is a significant energy cost for transferring even modest amounts of data between B/F nodes. Indeed, we can say that OptiLoop outperforms state-of-the-art alternatives {\em because} it properly accounts for the unique features of 5G, thus being more aggressive in deploying VNFs.

Comparing \Fig{details}(left) to \Fig{details}(center), we can see that OptiLoop deploys more VNFs than the optimum, but the number of VNFs per B/F node is similar. This is because OptiLoop activates slightly more B/F nodes than the optimum, as confirmed by \Fig{details}(right) showing that the average amount of traffic processed per B/F node is slightly lower in OptiLoop.

\subsection{Scaled-up network topology}

In the following, we investigate the performance of OptiLoop when used on larger-scale network topologies. To this end, based on indications from the mobile operator that provided us with the original topology described in \Sec{topo}, we generate a {\em scaled-up} version thereof. Specifically, we operate as follows:
\begin{enumerate}
    \item we replace each B/F node of the original topology with a ring of five B/S nodes;
    \item we place an additional 160~endpoints connected to 6,000~additional antennas;
    \item we connect each additional endpoint to two randomly-chosen B/S nodes;
    \item we set the traffic requested by the additional antennas in such a way that the traffic distribution matches the original one, scaled up by a factor of five.
\end{enumerate}
The resulting topology, depicted in \Fig{topo-big}, has over 200 B/F nodes serving traffic coming from 7,500 antennas. The results yielded by OptiLoop and the consolidation algorithm are reported in \Fig{scaledup}. Notice that there are no ``optimal'' curves, as computing the optimum for the scaled-up topology proved utterly impractical.

\Fig{scaledup}(left) shows that, as the topology gets larger, OptiLoop -- and, to a lesser extent, consolidation -- yield {\em more} savings, almost reaching~50\%. Intuitively, this is connected to the fact that in larger topologies it is easier to maintain connectivity while deactivating a substantial fraction of B/F nodes. This is confirmed by \Fig{scaledup}(center), showing that the spare CCAT, i.e., the unused computational power in the active network, is proportionally lower than in the original topology. Indeed, as we can see from \Fig{general}(center), the spare CCAT with the original topology reaches 2,500~units under OptiLoop, while in \Fig{scaledup}(center) it is below 10,000~units in spite of the topology being five times larger.

Finally, \Fig{scaledup}(right) breaks the total power consumption into its main components. By comparing it with \Fig{breakdown}, we can observe that:
\begin{itemize}
    \item the processing power is exactly five times larger than in the original topology, as that component is strictly proportional to the traffic to serve;
    \item the idle power is proportionally lower since, as observed earlier, there are fewer B/F nodes  activated only for sake of connectivity;
    \item the networking power is proportionally larger, as data are more likely to travel a longer path to the serving B/F node.
\end{itemize}
The latter two items suggest that networking power and idle power are, to a certain extent, antithetical, and it can be hard to minimize both at the same time.

\section{Conclusion}
\label{sec:conclusion}

We considered two of the unique features of 5G networks, namely, the hybrid nature of their nodes (which have both forwarding and computational capabilities) and the fact that the traffic to serve changes across processing steps. Such features require the entities in the MANO layer, and especially the NFVO, to make joint decisions about (i) which B/F nodes to activate, (ii) the VNF instances they run, and (iii) how to route traffic between VNFs and the nodes running them. We formulated a system model and optimization problem, that enable us to make all such decisions with the objective to minimize the energy consumption of the network. We further proposed OptiLoop, a solution concept based on integrating optimization within the MANO entities, allowing them to make decisions by repeatedly solving relaxed optimization problems.

We validated OptiLoop through a real-world testbed based on OpenDaylight and OpenStack, and further evaluated its performance through a large-scale emulated network whose topology and traffic are based on those of a major network operator. OptiLoop was shown to outperform state-of-the-art approaches and closely track the optimum, while representing only a minor contribution to the total network delay.

\bibliographystyle{IEEEtran}
\bibliography{refs}

\begin{thebibliography}{10}
\providecommand{\url}[1]{#1}
\csname url@samestyle\endcsname
\providecommand{\newblock}{\relax}
\providecommand{\bibinfo}[2]{#2}
\providecommand{\BIBentrySTDinterwordspacing}{\spaceskip=0pt\relax}
\providecommand{\BIBentryALTinterwordstretchfactor}{4}
\providecommand{\BIBentryALTinterwordspacing}{\spaceskip=\fontdimen2\font plus
\BIBentryALTinterwordstretchfactor\fontdimen3\font minus
  \fontdimen4\font\relax}
\providecommand{\BIBforeignlanguage}[2]{{%
\expandafter\ifx\csname l@#1\endcsname\relax
\typeout{** WARNING: IEEEtran.bst: No hyphenation pattern has been}%
\typeout{** loaded for the language `#1'. Using the pattern for}%
\typeout{** the default language instead.}%
\else
\language=\csname l@#1\endcsname
\fi
#2}}
\providecommand{\BIBdecl}{\relax}
\BIBdecl

\bibitem{swfan16_coopetition}
N.~Gazit, F.~Malandrino, and D.~Hay, ``{Coopetition between network operators
  and content providers in SDN/NFV core networks},'' in \emph{{IEEE INFOCOM
  SWFAN Workshop}}, 2016.

\bibitem{infocom15_optimal}
R.~Cohen, L.~Lewin-Eytan, J.~S. Naor, and D.~Raz, ``{Near optimal placement of
  virtual network functions},'' in \emph{{IEEE INFOCOM}}, 2015.

\bibitem{access16_joint}
L.~Wang, Z.~Lu, X.~Wen, R.~Knopp, and R.~Gupta, ``{Joint Optimization of
  Service Function Chaining and Resource Allocation in Network Function
  Virtualization},'' \emph{IEEE Access}, 2016.

\bibitem{infocom16_deploying}
T.-W. Kuo, B.-H. Liou, K.~C.-J. Lin, and M.-J. Tsai, ``{Deploying chains of
  virtual network functions: On the relation between link and server usage},''
  in \emph{{IEEE INFOCOM}}, 2016.

\bibitem{tc16_delayaware}
L.~Qu, C.~Assi, and K.~Shaban, ``Delay-aware scheduling and resource
  optimization with network function virtualization,'' \emph{IEEE Trans. on
  Communications}, 2016.

\bibitem{jsac17_buysell}
X.~Zhang, Z.~Huang, C.~Wu, Z.~Li, and F.~C. Lau, ``{An Online Stochastic
  Buy-Sell Mechanism for VNF chains in the NFV market},'' \emph{IEEE Journal on
  Selected Areas in Communications}, 2017.

\bibitem{cloudnet16_energyaware}
N.~El~Khoury, S.~Ayoubi, and C.~Assi, ``{Energy-Aware Placement and Scheduling
  of Network Traffic Flows with Deadlines on Virtual Network Functions},'' in
  \emph{IEEE CloudNet}, 2016.

\bibitem{epc_survey}
V.~G. Nguyen, A.~Brunstrom, K.~J. Grinnemo, and J.~Taheri, ``{SDN/NFV-based
  Mobile Packet Core Network Architectures: A Survey},'' \emph{IEEE
  Communications Surveys Tutorials}, 2017.

\bibitem{epc_split}
A.~Baumgartner, V.~S. Reddy, and T.~Bauschert, ``{Mobile core network
  virtualization: A model for combined virtual core network function placement
  and topology optimization},'' in \emph{IEEE NetSoft}, 2015.

\bibitem{globecom15_bearer}
G.~Hasegawa and M.~Murata, ``{Joint Bearer Aggregation and Control-Data Plane
  Separation in LTE EPC for Increasing M2M Communication Capacity},'' in
  \emph{IEEE GLOBECOM}, 2015.

\bibitem{qos_split}
S.~Khairi, M.~Bellafkih, and B.~Raouyane, ``{QoS management SDN-based for
  LTE/EPC with QoE evaluation: IMS use case},'' in \emph{SDS}, 2017.

\bibitem{vepc_archi}
X.~An, W.~Kiess, J.~Varga, J.~Prade, H.-J. Morper, and K.~Hoffmann,
  ``{SDN-based vs. software-only EPC gateways: A cost analysis},'' in
  \emph{IEEE NetSoft}, 2016.

\bibitem{tvt_mme}
J.~Prados-Garzon, J.~J. Ramos-Munoz, P.~Ameigeiras, P.~Andres-Maldonado, and
  J.~M. Lopez-Soler, ``{Modeling and Dimensioning of a Virtualized MME for 5G
  Mobile Networks},'' \emph{IEEE Trans. on Veh. Tech.}, 2017.

\bibitem{openair}
{OpenAirInterface: 5G software alliance for democratising wireless innovation}.
  \url{http://www.openairinterface.org}.

\bibitem{comsnets17_placement}
D.~Dietrich, C.~Papagianni, P.~Papadimitriou, and J.~S. Baras, ``{Network
  function placement on virtualized cellular cores},'' in \emph{COMSNETS},
  2017.

\bibitem{noi-wowmom18}
F.~Malandrino, C.~F. Chiasserini, C.~E. Casetti, and G.~Landi,
  ``{Optimization-in-the-Loop} for {Energy-Efficient} {5G},'' in \emph{IEEE
  WoWMoM}, 2018.

\bibitem{etsimano}
{ETSI}. (2017) {Network Functions Virtualisation (NFV); Management and
  Orchestration}.
  \url{http://www.etsi.org/deliver/etsi_gs/NFV-MAN/001_099/001/01.01.01_60/gs_NFV-MAN001v010101p.pdf}.

\bibitem{embedded}
J.~Mattingley and S.~Boyd, ``Cvxgen: A code generator for embedded convex
  optimization,'' \emph{Optimization and Engineering}, 2012.

\bibitem{boyd}
S.~Boyd and L.~Vandenberghe, \emph{Convex optimization}.\hskip 1em plus 0.5em
  minus 0.4em\relax Cambridge university press, 2004.

\bibitem{xhaul32}
{5G Crosshaul Project}. {Deliverable D3.2: Final XFE/XCI design and
  specification of southbound and northbound interfaces}.
  \url{http://5g-crosshaul.eu/wp-content/uploads/2018/01/5G-CROSSHAUL_D3.2.pdf
  }.

\bibitem{cisco}
Cisco, ``{Cisco Visual Networking Index},'' 2017.

\bibitem{lagopus-fun}
{Lagopus Project}. {It's kind of fun to do the impossible with DPDK}.
  \url{https://www.slideshare.net/lagopus/dpdk-summit-2015-its-kind-of-fun-to-do-the-impossible-with-dpdk}.

\end{thebibliography}

\end{document}